\documentclass{jpsj3_arXiv}
\usepackage{txfonts}

\usepackage[utf8]{inputenc}
\usepackage{physics}
\usepackage{amsmath, amssymb}
\usepackage{mathtools}
\usepackage{indentfirst}
\usepackage{graphicx}
\usepackage[colorlinks=true,linkcolor=blue,filecolor=blue,urlcolor=blue,citecolor=blue]{hyperref}

\usepackage{tikz}
\usetikzlibrary{positioning, arrows.meta, quotes}
\usepackage{bm}

\usepackage{accents}
\usepackage{cite}

\newcommand{\diff}[2]{\frac{d{#1}}{d{#2}}}
\newcommand{\pdiff}[2]{\frac{\partial{#1}}{\partial{#2}}}
\newcommand{\ppdiff}[3]{\frac{\partial^2{#1}}{\partial{#2} \partial{#3}}}
\newcommand{\oprho}{\hat{\rho}}
\newcommand{\opgamma}{\hat{\gamma}}

\newcommand{\whitedot}[1]{\accentset{\scriptscriptstyle\circ}{#1}}

\allowdisplaybreaks[1]

\counterwithin{equation}{section}

\title{Quantum Reversibility Meets Classical Reverse Diffusion}

\author{Ryota N{\sc asu}$^1$\thanks{nasu@wakayama-nct.ac.jp},
Gota T{\sc anaka}$^2$\thanks{gotanak@mi.meijigakuin.ac.jp}, and
Asato T{\sc suchiya}$^3$\thanks{tsuchiya.asato@shizuoka.ac.jp}}

\inst{$^1$Department of Electrical and Computer Engineering, National Institute of Technology, Wakayama College, 77 Noshima, Nada-Cho, Gobo, Wakayama 644-0023, Japan\\
$^2$Institute for Mathematical Informatics, Meiji Gakuin University, 1518 Kamikuratacho, Totsuka-ku, Yokohama 244-8539, Japan\\
$^3$Department of Physics, Shizuoka University, 836 Ohya, Suruga-ku, Shizuoka 422-8529, Japan}

\abst{
Bayes' rule connects forward and reverse processes in classical probability theory, and its quantum analogue has been discussed in terms of the Petz (transpose) map.
For
quantum dynamics governed by the Lindblad equation, the corresponding Petz map can also be written in Lindblad form.
In classical stochastic systems, the analogue of the Lindblad equation is the Fokker-Planck equation, and applying Bayes' rule to it yields the reverse diffusion equation underlying modern diffusion-based generative models.
It is known that a semiclassical approximation of the Lindblad equation yields
the Fokker-Planck equation for the Wigner function, which is
a quasiprobability distribution
defined on phase space as the Wigner transform of the density operator.
Here we demonstrate that
applying the same approximation to the Lindblad equation associated
with the Petz map produces an equation that coincides with that obtained from the Fokker-Planck equation via Bayes' rule.
This finding establishes a direct correspondence between the Petz map and Bayes' rule, unifying quantum reversibility
with classical reverse diffusion.}

\makeatletter
\long\def\@makefntext#1{\parindent 1em\noindent\hb@xt@1.8em{\hss\@makefnmark}#1}
\providecommand\ext@figure{lof}
\providecommand\ext@table{lot}
\providecommand{\newblock}{\hskip .11em\@plus.33em\@minus.07em}
\makeatother

\begin{document}
\maketitle

\section{Introduction}
Bayes' rule and Bayesian inference play a fundamental role in statistical reasoning
and have found wide applications across science.
They provide a natural framework for relating forward and reverse processes in classical probability theory.
It has been recognized that a formal quantum analogue of Bayes' rule, or its generalization known as Jeffrey's rule,
is given by the Petz (transpose) map \cite{Petz:1988usv}, which can be regarded
as a canonical reversal of quantum dynamics.
This correspondence has been further supported from the viewpoints of the fluctuation theorem \cite{Buscemi:2021oat} and the principle of minimum update \cite{Bai:2024xwv}.

When quantum dynamics, generally represented by a trace-preserving completely positive (TPCP) map, forms a semigroup,
it is governed by the Lindblad (or GKLS) equation \cite{Gorini:1975nb,Lindblad:1975ef}.
Previous studies \cite{kwon2019fluctuation,Kwon:2021itt} have shown that the Petz map corresponding to a Lindblad
dynamics can itself be expressed in the Lindblad form.
In classical stochastic systems, the analogue of the Lindblad equation is the Fokker–Planck equation, which describes
the time evolution of probability distributions as a diffusion process.
It has also been shown \cite{frigerio1984diffusion,tzanakis1998generalized,strunz1998classical,dubois2021semi,Hernandez:2023yxf,Yoneya:2024asd,Yoneya:2025wel} that,
when the quantum system can be considered as close to
a classical one, the Lindblad equation reduces to the Fokker–Planck equation.
Applying Bayes' rule to the Fokker-Planck equation
yields the reverse diffusion equation \cite{Anderson1982}, which underlies modern diffusion-based generative models \cite{pmlr-v37-sohl-dickstein15}.

In this work, we establish a more direct correspondence between Bayes' rule and the Petz map.
We consider a semiclassical approximation of the Lindblad equation \cite{Hernandez:2023yxf}
that reduces to the Fokker–Planck equation for the Wigner function, which is
a quasiprobability distribution defined on phase
space as the Wigner transform of the density operator.
We show that applying the same approximation to the Lindblad equation associated with the Petz map yields another
equation for the Wigner function that coincides with that obtained from the Fokker–Planck equation via Bayes' rule.
This finding reveals a direct correspondence between the Petz map and Bayes' rule, highlighting a unified view of quantum reversibility and classical reverse diffusion.
The above mentioned relationship is schematically
summarized in Fig. 1.

This paper is organized as follows.
In section 2, we review the Petz map and
the Lindblad equation. We see that the Petz map for
the Lindblad dynamics takes the Lindblad form.
We also review
the Fokker-Planck equation and
its reverse diffusion process that is obtained by applying
Bayes' rule to the former.
In section 3, we first see how a semiclassical approximation
reduces
the Lindblad equation to the Fokker-Planck equation
for the Wigner function,
and then show that the same approximation
of the Lindblad equation with the Petz map yields
a diffusion equation that is
obtained by applying Bayes' rule
to the Fokker-Planck equation.
In section 4, we construct WKB solutions of the semi-classical Lindblad equation
and its reversal semi-classical equation.
Section 5 is devoted to conclusion and outlook.
In appendix A, details of calculations are gathered.
In appendix B, an example of the WKB analysis is given.

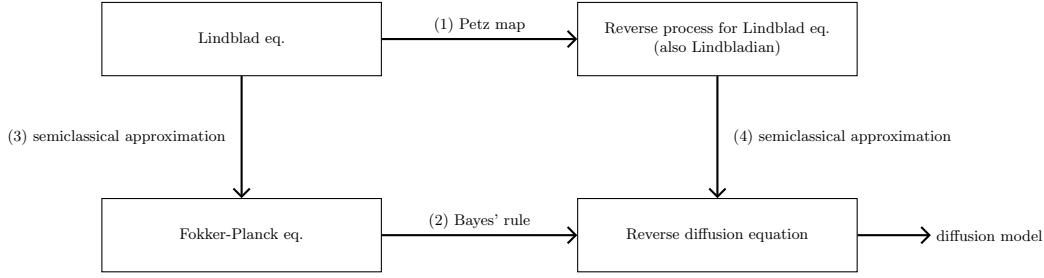
\begin{figure*}[t]
    \centering
    \resizebox{\textwidth}{!}{%
    \begin{tikzpicture}[
        >=Stealth,
        box/.style={rectangle, draw, text width=6cm, align=center, minimum height=1.5cm},
        arrow/.style={-Straight Barb, thick}]
    \node[box] (lindblad) {Lindblad eq.};
    \node[box, below=2.5cm of lindblad] (convection) {Fokker-Planck eq.};
    \node[box, right=4cm of lindblad] (reversal) {Reverse process for Lindblad eq. \\ (also Lindbladian)};
    \node[box] (reverse_time) at (reversal |- convection) {Reverse diffusion equation};
    \node[right=1.5cm of reverse_time, align=center] (sbdiff) {diffusion model};
    \draw[arrow] (lindblad) -- (convection) node[midway, left=0.2cm, align=center] {(3) semiclassical approximation};
    \draw[arrow] (lindblad) to["(1) Petz map"] (reversal);
    \draw[arrow] (convection) to["(2) Bayes' rule"] (reverse_time);
    \draw[arrow] (reversal) -- (reverse_time) node[midway, right=0.2cm, align=center] {(4) semiclassical approximation};
    \draw[arrow] (reverse_time) -- (sbdiff);
    \end{tikzpicture}%
    }
    \caption{Schematic relationship between the Lindblad and Fokker-Planck equations. The Petz map provides a reversal process for the Lindblad equation [arrow (1)], while Bayes' rule plays the analogous role for the Fokker-Planck equation [arrow (2)]. A semiclassical approximation [arrows (3) and (4)] connects the quantum and classical descriptions, showing the correspondence between the Petz map and Bayes' rule.}
    \label{fig:summary}
\end{figure*}

\section{Reviews}
In this section, we give brief reviews on the Petz map and the Lindblad equation (section 2.1) and on
the Fokker-Planck equation and reverse diffusion process (section 2.2).
\subsection{Petz map and Lindblad equation}
The Petz (transpose) map was originally introduced in Ref.~\cite{Petz:1988usv} as the equality condition for the data-processing inequality of the quantum relative entropy, and later rediscovered in the contexts of quantum error correction \cite{Barnum:2002bfd} and statistical physics \cite{PhysRevA.77.034101}. It provides a concrete formulation of a reversal process for quantum channels and has been discussed as a quantum analogue of Bayes’ rule.
For a TPCP map $\mathcal{A}$ and
a reference state $\hat{\gamma}$, the Petz map is defined by
\begin{align*}
    \mathcal{P}_{\mathcal{A},\hat{\gamma}}(\cdot) \coloneqq \hat{\gamma}^{1/2}\mathcal{A}^{\dagger}\left(\mathcal{A}(\hat{\gamma})^{-1/2} (\cdot)\mathcal{A}(\hat{\gamma})^{-1/2}\right)\hat{\gamma}^{1/2} \ ,
\end{align*}
where $\mathcal{A}^\dagger$ denotes the adjoint map with respect to the Hilbert–Schmidt inner product.
This map defines a TPCP operation satisfying
   $ \mathcal{P}_{\mathcal{A},\hat{\gamma}} \circ \mathcal{A}(\hat{\gamma}) = \hat{\gamma}$.
When the data-processing inequality is saturated, the Petz map exactly restores the original state after the channel action, providing a canonical construction of quantum reversibility.

The Lindblad equation describes the most general form of Markovian quantum dynamics as a one-parameter semigroup of TPCP maps,
\begin{equation}
\frac{\partial \hat{\rho}_t}{\partial t} = \mathcal{L}_t(\hat{\rho}_t) \ ,
\label{eq:lindblad}
\end{equation}
with the generator
\begin{align}
    \mathcal{L}_{t}(\hat{\rho}_t) &= -\frac{i}{\hbar}\left[\hat{H}_t,\hat{\rho}_t\right]_{-}
    +\sum_{\alpha}\left(\hat{L}_{\alpha,t}\hat{\rho}_{t}\hat{L}_{\alpha,t}^{\dagger}-\frac{1}{2}\left[\hat{L}_{\alpha,t}^{\dagger}\hat{L}_{\alpha,t}, \hat{\rho}_{t}\right]_{+}\right) \ . \nonumber
\end{align}
Here
$\hat{\rho}_t$ is the density operator,
$\hat{H}_t$ is a Hermitian Hamiltonian,
and $\hat{L}_{\alpha,t}$ are Lindblad operators that are not necessarily
Hermitian.

The Lindblad equation generates a TPCP map
$\mathcal{T}_{\Delta t}=(1+\Delta t \mathcal{L}_t)$, from which the corresponding Petz map $\mathcal{P}_{\mathcal{T}_{\Delta t},\hat{\gamma}_{t-\Delta t}}$ interpreted as the reversal of the dynamics can be constructed\cite{footnotegammat}.
It has been shown in Refs. \cite{kwon2019fluctuation, Kwon:2021itt} that this Petz map leads to a reversed Lindblad equation\cite{footnoteddt},
\begin{align}
        -\pdv{\hat{\rho}_{t}}{t} &= -\frac{i}{\hbar}\left[\hat{\tilde{H}}_t,\oprho_{t}\right]_{-}
        + \sum_\alpha \left( \hat{\tilde{L}}_{\alpha,t} \oprho_{t} \hat{\tilde{L}}_{\alpha,t}^{\dagger} - \frac{1}{2} \left[ \hat{\tilde{L}}_{\alpha,t}^{\dagger} \hat{\tilde{L}}_{\alpha,t}, \oprho_{t} \right]_{+}\right)   \ ,
        \label{eq:reversal Lindblad eq}
\end{align}
with the corresponding reversed operators
\begin{align}
    \hat{\tilde{H}}_t = -\frac{1}{2}&\left(\hat{G}_t\hat{H}_t\hat{G}_t^{-1} + i\hbar \dot{\hat{G}}_t\hat{G}_t^{-1}
        + \frac{i\hbar}{2}\sum_\alpha\hat{G}_t\hat{L}_{\alpha,t}^{\dagger}\hat{L}_{\alpha,t}\hat{G}_t^{-1}\right) + h.c.   \ , \label{eq:reversal hamiltonian}
\end{align}
\begin{align}
    &\hat{\tilde{L}}_{\alpha,t} = \hat{G}_t\hat{L}_{\alpha,t}^{\dagger}\hat{G}_t^{-1}  \ ,  \label{eq:reversal Lindblad operator}
\end{align}
where
\begin{align}
\hat{G}_t = \opgamma_t^{1/2},  \;\; \dot{\hat{G}}_t = \frac{d \hat{G}_t}{dt} \ . \label{G}
\end{align}
Thus, the Petz map associated with a Lindblad dynamics can itself be
expressed in Lindblad form,
establishing a consistent framework for describing quantum reversibility
within the Markovian regime.

\subsection{Fokker-Planck equation and reverse diffusion process}
We review the Fokker–Planck equation for classical stochastic dynamics and recall that applying Bayes’ rule yields the reverse diffusion equation \cite{Anderson1982} underlying modern diffusion-based generative models.

We consider a classical system with $n$ degrees of freedom parameterized by
$\boldsymbol{x}=(x^1,x^2,\ldots,x^n)$.
The stochastic process is described by the Langevin equation
\begin{align}
    \diff{x^\mu(t)}{t} = f^{\mu}(\bm{x},t) + &\sum_{\nu=1}^n {g^{\mu}}_{\nu}(\bm{x},t)\eta^{\nu}(t)
     \quad (\mu=1,2,\ldots,n)\ , \label{eq:Langevin eq}
\end{align}
where $\eta^{\mu}(t)$ denotes Gaussian white noise satisfying $\expval{\eta^{\mu}(t)\eta^{\nu}(t')} = \delta^{\mu\nu}\delta(t-t')$ and $f^\mu$ represent the drift terms.
The probabilistic distribution $P(\bm{x},t)$
follows the Fokker-Planck equation
\begin{align}
    \pdv{P(\bm{x},t)}{t} = - \sum_\mu \pdv{x^\mu}
    \left(f^{\mu}(\bm{x},t)P(\bm{x},t) \right) + \frac{1}{2}\sum_{\mu,\nu}\ppdiff{}{x^\mu}{x^\nu}\left(G^{\mu\nu}(\bm{x},t)P(\bm{x},t)\right) \ , \label{eq:Fokker-Planck eq}
\end{align}
where $G^{\mu\nu}=\sum_\lambda {g^\mu}_\lambda {g^\nu}_\lambda$.

The backward Fokker-Planck equation for the conditional probability $P(\bm{x}_s,s|\bm{x}_t,t)$ where $s \geq t$ is also obtained from Eq. \eqref{eq:Langevin eq},
\begin{align}
    -\pdv{P(\bm{x}_s,s|\bm{x}_t,t)}{t} &= \sum_\mu f^{\mu}(\bm{x}_t,t)\pdv{P(\bm{x}_s,s|\bm{x}_t,t)}{x^\mu_t}
        + \frac{1}{2}\sum_{\mu,\nu}G^{\mu\nu}(\bm{x}_t,t)\pdv{P(\bm{x}_s,s|\bm{x}_t,t)}{x_t^\mu}{x_t^\nu} \ . \label{eq:backward Fokker-Planck eq}
\end{align}
Bayes' rule tells that the conditional probability $P(\bm{x}_t,t|\bm{x}_s,s)$
is expressed as
\begin{align}
    P(\bm{x}_t,t|\bm{x}_s,s) = \frac{P(\bm{x}_s,s|\bm{x}_t,t)P(\bm{x}_t,t)}{P(\bm{x}_s,s)}
    \ . \label{eq:Bayes' theorem for transition probability}
\end{align}
We introduce a probability distribution $\tau(\bm{x}_s)$ at time $s$
and define a probability distribution for reverse process $\bar{P}(\bm{x},t)$
by
$\bar{P}(\bm{x}_t,t)= \int d^d\bm{x}_s P(\bm{x}_t,t|\bm{x}_s,s)\tau(\bm{x}_s)$ .
By using Eqs. \eqref{eq:Fokker-Planck eq},
\eqref{eq:backward Fokker-Planck eq} and
\eqref{eq:Bayes' theorem for transition probability},
we obtain the reverse-time diffusion equation,
\begin{align}
    -\pdiff{\bar{P}(\bm{x}_t,t)}{t} &= -\sum_\mu \pdiff{}{x_t^\mu}\left(\bar{f}^\mu(\bm{x}_t,t)\bar{P}(\bm{x}_t,t)\right)
        + \frac{1}{2}\sum_{\mu,\nu}\ppdiff{}{x_t^\mu}{x_t^\nu}\left(G^{\mu\nu}(\bm{x}_t,t)\bar{P}(\bm{x}_t,t)\right) \ , \label{eq: reverse time diffusion equation}
\end{align}
where the reversed drift terms are
\begin{align*}
    \bar{f}^{\mu}(\bm{x}_t,t) &= -f^\mu(\bm{x}_t,t)
        + \frac{1}{P(\bm{x}_t,t)}\sum_\nu\pdiff{}{x_t^\nu}\left(G^{\mu\nu}(\bm{x}_t,t)P(\bm{x}_t,t)\right) \ .
\end{align*}
A notable feature of the reverse-time diffusion equation is that its drift term encodes information of the forward process and drives the distribution toward the initial state $P(\bm{x},0)$.
This provides the theoretical foundation for score-based
diffusion models \cite{pmlr-v37-sohl-dickstein15, song2019generative, song2020score}.

\section{Semiclassical approximation of Lindblad and reversal
Lindblad equations}

First,
we see that the Fokker–Planck equation for the Wigner
function can be derived from the Lindblad equation via a
semiclassical approximation based on an expansion in $\hbar$
\cite{frigerio1984diffusion,tzanakis1998generalized,strunz1998classical,dubois2021semi,Hernandez:2023yxf}.

We consider a quantum system with
$N$ degrees of freedom, where all operators are constructed
in terms of $\hat{q}_i$ and $\hat{p}_i$ $(i=1,\ldots,N)$, which obey
the commutation relations
\begin{align*}
[\hat{q}_i,\hat{q}_j]_-=0 \ , \;\; [\hat{p}_i,\hat{p}_j]_-=0 \ , \;\;
[\hat{q}_i,\hat{p}_j]_-=i\hbar \delta_{ij} \ .
\end{align*}
The states $|\boldsymbol{q}\rangle$, defined by
$\hat{q}_i|\boldsymbol{q}\rangle=q_i|\boldsymbol{q}\rangle $ with $\boldsymbol{q}=
(q_1,q_2,\ldots,q_N)$, form a basis of the Hilbert space.
We also consider the corresponding (semi)classical system, whose
phase space is denoted by
$(\boldsymbol{Q},\boldsymbol{P})$.

We
introduce the Wigner transformation, which maps operators on
the Hilbert space to functions on phase space and is defined as
\begin{align*}
    O(\bm{Q},\bm{P}) &= \int d^N \sigma \bra{\bm{Q}+\frac{\bm{\sigma}}{2}}\hat{O}(\hat{\boldsymbol{q}},\hat{\boldsymbol{p}})\ket{\bm{Q}-\frac{\bm{\sigma}}{2}}e^{-\frac{i}{\hbar}\bm{P}\cdot\bm{\sigma}} \ ,
\end{align*}
where $\bm{Q}$ and $\bm{\sigma}$ are the center of mass and relative coordinates, respectively,
defined by\cite{footnoteinverseWigner}
\begin{align*}
    \bm{Q}=\frac{\bm{q}_1 + \bm{q}_2}{2}, \quad \bm{\sigma} = \bm{q}_1 - \bm{q}_2 \  .
\end{align*}

The Wigner transformation satisfies the following relation
\begin{align}
    \hat{O}_1\hat{O}_2 \xrightarrow[]{{\text{Wigner}}} &O_1(\bm{Q},\bm{P}) \star O_2(\bm{Q},\bm{P}) \nonumber   \\
        &\coloneqq \exp\left[\frac{i\hbar}{2}\sum_{i=1}^{N}\left(\pdiff{}{Q_i}\pdiff{}{P'_i} - \pdiff{}{Q'_i}\pdiff{} {P_i}\right)\right]
            \left.\vphantom{\pdiff{}{Q'_i}}\mathcal{O}_1(\bm{Q},\bm{P})\mathcal{O}_2(\bm{Q'},\bm{P'})\right|_{\substack{\bm{Q'}=\bm{Q} \\ \bm{P'} = \bm{P}}}
    \label{Moyal product}
\end{align}
Expanding the exponential yields
\begin{align}
    &\mathcal{O}_1\mathcal{O}_2 + \frac{i\hbar}{2}\left\{\mathcal{O}_1,\mathcal{O}_2\right\}_p
    - \frac{\hbar^2}{8}\sum_i\left[\left\{\pdiff{\mathcal{O}_1}{Q_i},\pdiff{\mathcal{O}_2}{P_i}\right\}_p - \left\{\pdiff{\mathcal{O}_1}{P_i},\pdiff{\mathcal{O}_2}{Q_i}\right\}_p\right] + \order{\hbar^3} \ . \label{hbarexpansion}
\end{align}
Here, $\{,\}_p$ denotes the Poisson brackets, and $\star$
represents the star (Moyal) product.
The trace of an operator can be written as an integral over phase space,
\begin{align}
   \mbox{Tr} (\hat{\mathcal{O}})= \int \frac{d^N Q \, d^N P}{(2\pi\hbar)^N}
   O(\bm{Q}, \bm{P})  \ .  \label{trace}
\end{align}

In particular, the Wigner transform of the density operator
$\hat{\rho}$ defines the Wigner function, denoted by
$W^{(\rho)}(\bm{Q},\bm{P})$.
Because $\hat{\rho}$ is Hermitian, $W_{t}^{(\rho)}$ is real.
Using Eqs. \eqref{Moyal product} and \eqref{trace},
     the expectation value of an observable $\hat{O}(\hat{q}, \hat{p})$ is obtained as
  \begin{align*}
    \langle \hat{O} \rangle = \Tr[\hat{\rho} \hat{O}]
    = \int \frac{d^N Q \, d^N P}{(2\pi\hbar)^N} \, O(\bm{Q}, \bm{P}) \, W_{t}^{(\rho)}(\bm{Q}, \bm{P}) \ .
  \end{align*}
These are the properties required of a probability distribution on
phase space.
However, since the Wigner function is not guaranteed to be positive
semi-definite, it is generally regarded as a quasiprobability
distribution.

We now examine the Wigner transformation of the Lindblad equation
\eqref{eq:lindblad}.
We denote the Weyl transforms of
the Hamiltonian $\hat{H}_t$ and
the Lindblad operators $\hat{L}_{\alpha,t}$ by
$H_t(\bm{Q},\bm{P})$ and $\ell_{\alpha,t}(\bm{Q},\bm{P})/\sqrt{\hbar}$, respectively, and assume that $H_t(\bm{Q},\bm{P})$ and
$\ell_{\alpha,t}(\bm{Q},\bm{P})$ do not include
$\hbar$.
We then introduce the $2N$-dimensional phase-space vector
$\bm{x}=(\bm{Q},\bm{P})$, where
$x^\mu=Q_{\mu}$ for $\mu=1,2,\ldots,N$ and $x^\mu=P_{\mu-N}$ for $\mu=N+1,N+2,\ldots,2N$.
We also define a symplectic form $\omega^{\mu\nu}$ by
$\omega^{i\:i+N}=1$, $\omega^{i+N \: i}=-1$ for $i=1\ldots, N$ and
other $\omega^{\mu\nu}$'s $=0$.
Under these assumptions,
expanding the Wigner transform of the Lindblad equation \eqref{eq:lindblad} up to $\mathcal{O}(\hbar)$ using Eqs. \eqref{Moyal product} and
\eqref{hbarexpansion} yields
\begin{align}
    \pdiff{W_{t}^{(\rho)}}{t}&=-\sum_{\mu}\pdv{}{x^{\mu}}\left( f^{\mu}(\bm{x},t) W_{t}^{(\rho)} \right)
    + \frac{1}{2}\sum_{\mu,\nu} \frac{\partial^2}{\partial x^{\mu} \partial x^{\nu}} \left( G^{\mu\nu}(\bm{x},t) W_{t}^{(\rho)} \right) \ , \label{eq:Wigner tran of Lindblad eq}
\end{align}
where $f^{\mu}$ and $G^{\mu\nu}$ are the drift and diffusion
coefficients, respectively, defined by
\begin{align}
    f^{\mu}(\bm{x},t) = &\sum_\nu \omega^{\mu\nu}
        \left[ \pdv{H_t}{x^\nu}
        + \sum_\alpha \left( \mathrm{Im} \left( \ell_{\alpha,t} \frac{\partial \ell_{\alpha,t}^*}{\partial x^\nu} \right)
         - \frac{\hbar}{2}\mathrm{Re}\left\{ \ell_{\alpha,t}, \frac{\partial \ell_{\alpha,t}^*}{\partial x^\nu} \right\}_p \right) \right] \label{drift}
\end{align}
and
\begin{align}
G^{\mu\nu}(\bm{x},t) =\hbar \sum_{\lambda,\rho} \sum_\alpha \omega^{\mu\lambda}\omega^{\nu\rho}
\mbox{Re} \left( \dfrac{\partial \ell_{\alpha,t}}{\partial x^\lambda }\dfrac{\partial \ell_{\alpha,t}^*}{\partial x^\rho}  \right) \ .
\label{diffusion coefficient}
\end{align}
Details of the derivation are provided in appendix A.1.
Consequently,
Eq. \eqref{eq:Wigner tran of Lindblad eq} takes the form
of the Fokker-Planck equation \eqref{eq:Fokker-Planck eq}.
Moreover,
it can be shown that
$G^{\mu\nu}$
is positive semi-definite\cite{footnoteGmunu}
, ensuring that the evolution preserves the probabilistic interpretation of $W_{t}^{(\rho)}$\cite{footnoteLindbladeq}
.
Thus, we have seen that the Lindblad equation reduces to the Fokker–Planck equation for the Wigner function $W_{t}^{(\rho)}$
under the semiclassical approximation.

We finally examine the semiclassical approximation of the reversed Lindblad equation \eqref{eq:reversal Lindblad eq}.
The Wigner function of the reference state $\hat{\gamma}$
is denoted by $W^{(\gamma)}_t$.

By applying the same procedure above,
we obtain an equation for the Wigner function $W^{(\rho)}_t$
\begin{align}
    -\pdv{W_{t}^{(\rho)}}{t} &= -\sum_{\mu}\pdv{}{x^{\mu}}\left( \tilde{f}^{\mu}(\bm{x},t) W_{t}^{(\rho)} \right)
    + \frac{1}{2}\sum_{\mu,\nu} \frac{\partial^2}{\partial x^{\mu} \partial x^{\nu}} \left( \tilde{G}^{\mu\nu}(\bm{x},t) W_{t}^{(\rho)} \right) \ ,\label{Wigner tran of inverse Lindblad eq}
\end{align}
where $\tilde{f}^{\mu}$ and $\tilde{G}^{\mu\nu}$ are defined by
\begin{align}
    \tilde{f}^{\mu}(\bm{x},t) = &\sum_\nu \omega^{\mu\nu}
        \left[ \pdv{\tilde{H}_t}{x^\nu}
        + \sum_\alpha \left( \mathrm{Im} \left( \tilde{\ell}_{\alpha,t} \frac{\partial \tilde{\ell}_{\alpha,t}^*}{\partial x^\nu} \right)
            - \frac{\hbar}{2}\mathrm{Re}\left\{ \tilde{\ell}_{\alpha,t}, \frac{\partial \tilde{\ell}_{\alpha,t}^*}{\partial x^\nu} \right\}_p \right) \right] \ , \label{drift 1}
\end{align}
\begin{align}
\tilde{G}^{\mu\nu}(\bm{x},t) =\hbar \sum_{\lambda,\rho} \sum_\alpha \omega^{\mu\lambda}\omega^{\nu\rho}
\mbox{Re} \left( \dfrac{\partial \tilde{\ell}_{\alpha,t}}{\partial x^\lambda }\dfrac{\partial \tilde{\ell}_{\alpha,t}^*}{\partial x^\rho}  \right) \ . \nonumber
\end{align}
Here, $\tilde{H}_t$ and $\tilde{\ell}_{k,t}$ still contain the star products inside.
Expanding these star products up to $\mathcal{O}(\hbar)$
reduces Eq. \eqref{Wigner tran of inverse Lindblad eq} to
\begin{align}
    -\pdv{W_{t}^{(\rho)}}{t} &= -\sum_{\mu}\pdv{}{x^{\mu}}\left( \bar{f}^{\mu}(\bm{x},t) W_{t}^{(\rho)} \right)
    + \frac{1}{2}\sum_{\mu,\nu} \frac{\partial^2}{\partial x^{\mu} \partial x^{\nu}} \left( G^{\mu\nu}(\bm{x},t) W_{t}^{(\rho)} \right) \ , \label{eq:1st order of Wigner tran of inverse Lindblad eq}
\end{align}
with
\begin{align}
    \bar{f}^{\mu}(\bm{x},t) =& -f^{\mu}(\bm{x},t)
    + \frac{1}{W_{t}^{(\gamma)}}\sum_{\nu}\pdv{}{x^{\nu}}\qty(G^{\mu\nu}(\bm{x},t)W_{t}^{(\gamma)}) \ . \nonumber
\end{align}
Here $f^{\mu}$ and $G^{\mu\nu}$ are defined in
Eqs. \eqref{drift}
and \eqref{diffusion coefficient}, respectively.
Details of the derivation are provided in appendix A.2.

Eq. \eqref{eq:1st order of Wigner tran of inverse Lindblad eq}
takes the form of the reverse-time diffusion
equation \eqref{eq: reverse time diffusion equation}.
Thus, we have shown that the reversed Lindblad equation reduces to the reverse-time diffusion equation under the semiclassical approximation.
The former corresponds to the Petz map associated with the Lindblad
equation, while the latter is obtained by applying Bayes’ rule to the
Fokker–Planck equation, which represents its semiclassical
approximation.
This correspondence provides a new insight into the connection between quantum reversibility and classical reverse diffusion, linking
the Petz map with Bayes’ rule.

\section{Construction of WKB solutions}
In this section, we construct WKB solutions for
the semi-classical Lindblad equation \eqref{eq:Wigner tran of Lindblad eq}
(section 4.1) and its reversal
semi-classical equation
\eqref{eq:1st order of Wigner tran of inverse Lindblad eq} (section 4.2)
\cite{footnoteWKB}
.
We examine a concrete example in appendix B.
For simplicity, we consider the $N=1$ case.
Generalization to general $N$ cases
is straightforward. First, we define $K_{\mu}$ and $J_{\mu}$ as
\begin{align}
    K_{\mu}(x,t) &= \sum_{\alpha}\Im{\ell_{\alpha,t}\pdv{\ell_{\alpha,t}^*}{x^\mu}} \ ,\\
    J_{\mu}(x,t) &= \sum_{\alpha}\Re\qty{\ell_{\alpha,t},\pdv{\ell_{\alpha,t}^*}{x^{\mu}}}_{p} \ ,
\end{align}
and express the drift term in Eq. \eqref{eq:Wigner tran of Lindblad eq} as
\begin{align}
    f^{\mu}(x,t) = \sum_{\mu}\omega^{\mu\nu}\qty(\pdv{H_t}{x^\nu} + K_{\nu}(x,t) - \frac{\hbar}{2}J_{\nu}(x,t)) \ .
\end{align}

\subsection{Solution for the forward process.}
Here, we assume that the Wigner function $W_t^{(\rho)}(x)$ can be expanded in terms of $\hbar$ as
\begin{align}
    W^{(\rho)}_t = W_{t,0}^{(\rho)} + \hbar W_{t,1}^{(\rho)} + \cdots \ .\label{eq:hbar expansion for forward Wigner function}
\end{align}
By substituting Eq. \eqref{eq:hbar expansion for forward Wigner function} into Eq. \eqref{eq:Wigner tran of Lindblad eq} and organizing
terms order by order in $\hbar$, we obtain
\begin{align}
    \pdv{W_{t,0}^{(\rho)}}{t} &= -\sum_{\mu,\nu}\pdv{x^{\mu}}\qty[\omega^{\mu\nu}\qty(\pdv{H_t}{x^{\nu}} + K_{\nu}(x,t))W_{t,0}^{(\rho)}]\ , \label{eq:0th order equation in forward process}\\
    \pdv{W_{t,1}^{(\rho)}}{t} &= -\sum_{\mu,\nu}\pdv{x^{\mu}}\qty[\omega^{\mu\nu}\qty(\pdv{H_t}{x^{\nu}} + K_{\nu}(x,t))W_{t,1}^{(\rho)}] \nonumber \\
    &\qquad +\frac{1}{2}\qty[\pdv{x^{\mu}}\qty(\omega^{\mu\nu}J_{\nu}(x,t)W_{t,0}^{(\rho)})+\pdv{}{x^\mu}{x^\nu}\qty(D^{\mu\nu}(x,t)W_{t,0}^{(\rho)})] \  , \label{eq:1st order equation in forward process}
\end{align}
where $D^{\mu\nu}=G^{\mu\nu}/\hbar$.
Further calculations give
\begin{align}
    &\pdv{W_{t,0}^{(\rho)}}{t} + A(x,t)\pdv{W_{t,0}^{(\rho)}}{Q} + B(x,t)\pdv{W_{t,0}^{(\rho)}}{P} + \Gamma(x,t)W_{t,0}^{(\rho)} = 0 \ ,\label{eq:0th order equation in forward process 2}\\
    &\pdv{W_{t,1}^{(\rho)}}{t} + A(x,t)\pdv{W_{t,1}^{(\rho)}}{Q} + B(x,t)\pdv{W_{t,1}^{(\rho)}}{P} + \Gamma(x,t)W_{t,1}^{(\rho)} = \Delta_0(x,t) \ , \label{eq:1st order equation in forward process 2}
\end{align}
where we defined
\begin{align}
    A(x,t) &= \pdv{H_t}{P} + K_{P}(x,t) \ , \\
    B(x,t) &=  -\pdv{H_t}{Q} - K_{Q}(x,t) \ , \\
    \Gamma(x,t) &= \omega^{\mu\nu}\pdv{K_{\nu}(\bm{x},t)}{x^{\mu}} = \pdv{K_{P}(x,t)}{Q} - \pdv{K_{Q}(x,t)}{P} \ , \\
    \Delta_0(x,t) &= \frac{1}{2}\qty[\pdv{x^{\nu}}\qty(\omega^{\mu\nu}J_{\nu}(x,t)W_0^{(\rho)}) + \ppdiff{}{x^\mu}{x^\nu}\qty(D^{\mu\nu}(x,t)W_0^{(\rho)})] \ .
\end{align}

Since $\Delta_0(x(t),t)$ is determined once $W_{t,0}^{(\rho)}$ is known, we first solve Eq. \eqref{eq:0th order equation in forward process 2}. In what follows, we denote the total time derivative by a dot, such as $\dot{f}$.
Eq. \eqref{eq:0th order equation in forward process 2} can be solved by using the method of characteristics. Namely, introducing characteristic curves $x(t) = (Q(t),\,P(t))$ satisfying the characteristic equations
\begin{align}
    \dot{Q}(t) = A(x(t),t) \ ,\quad  \dot{P}(t) = B(x(t),t) \ , \label{eq:characteristic_eq}
\end{align}
we have
\begin{align}
    \dot{W}^{(\rho)}_{t,0}(x(t))
    = \left.\pdv{W_{t,0}^{(\rho)}}{t}\right|_{x=x(t)} + \left.A(x(t),t)\pdv{W_{t,0}^{(\rho)}}{Q}\right|_{x=x(t)} + \left.B(x(t),t)\pdv{W_{t,0}^{(\rho)}}{P}\right|_{x=x(t)} \ .
\end{align}
On the characteristic curve, therefore, we obtain
\begin{align}
    \dot{W}_{t,0}^{(\rho)}(x(t)) + \Gamma(x(t),t)W_{t,0}^{(\rho)}(x(t)) = 0 \ .
\end{align}
Solving this, we find along the characteristic curve
\begin{align}
    W_{t,0}^{(\rho)}(x(t)) = W_{0,0}^{(\rho)}(x_0) \exp[-\int_0^t ds \: \Gamma(x(s),s)] \ .\label{eq:solution_for_W0_eq_on_characteristic_curve}
\end{align}
Here $x_0 = (Q_0,\,P_0)$ are the initial conditions of the characteristic curve, namely $Q(0)=Q_0,\,P(0)=P_0$. The characteristic equation \eqref{eq:characteristic_eq} can be regarded as a deformation of Hamilton's equations of motion by the environmental contribution $K_\mu$.
From Eq. \eqref{eq:solution_for_W0_eq_on_characteristic_curve},
we now construct the general solution to
Eq. \eqref{eq:0th order equation in forward process 2}
depending on the three independent variables $t$, $x=(Q,P)$.
We regard the characteristic curve as a map (including
a one-parameter family) that sends the initial values $x_0 = (Q_0,\,P_0)$ to $x(t) = (Q(t),\,P(t))$, and write
\begin{align}
    x(t) = \Phi_t(x_0)  \ .
\label{Phi}
\end{align}
Then, at time $t$, the initial values of the characteristic curve satisfying $x(t) \coloneqq (Q(t),P(t))= (Q,P) \eqqcolon x$ are given by
\begin{align}
    x_0 = \Phi^{-1}_t(x) \ .
\label{Phi_inverse}
\end{align}
Hence, from
Eq. \eqref{eq:solution_for_W0_eq_on_characteristic_curve},
the general solution to
Eq. \eqref{eq:0th order equation in forward process 2} is found to be
\begin{align}
    W_{t,0}^{(\rho)}(x) = W_{0,0}^{(\rho)}(\Phi_t^{-1}(x)) \exp\qty[\int_0^{t}ds  \:
    \Gamma(\Phi_{s}\circ\Phi_t^{-1}(x),s)] \ .\label{eq:solution_for_W0_eq}
\end{align}
This solution contains one arbitrary function, namely the initial distribution $W_0^{(\rho)}(0,x)$.

Once $W_{t,0}^{(\rho)}$ has been determined, we can calculate
$\Delta_0(x,t)$ from it, and then solve Eq. \eqref{eq:1st order equation in forward process 2}. Eq. \eqref{eq:1st order equation in forward process 2} can also be solved by using
the method of characteristics. The characteristic equations are again given by Eq. \eqref{eq:characteristic_eq}, just as in the case of Eq. \eqref{eq:0th order equation in forward process 2}. Then, along the characteristic curve, we obtain
\begin{align}
    \dot{W}_{t,1}^{(\rho)}(x(t)) + \Gamma(x(t),t) W_{t,1}^{(\rho)}(x(t)) = \Delta_0(x(t),t) \ .
\end{align}
Now we define
\begin{align}
    \tilde{W}_{t,1}^{(\rho)}(x(t)) = W_{t,1}^{(\rho)}(x(t))\exp\qty[\int_0^t ds \Gamma(x(s),s)] \ .
\end{align}
Then, we have
\begin{align}
    \dot{\tilde{W}}_{t,1}^{(\rho)}(x(t))
    &= \dot{W}_{t,1}^{(\rho)}(x(t))\exp\qty[\int_0^t ds \Gamma(x(s),s)] \nonumber\\
    &\qquad+\Gamma(x(t),t)W_{t,1}^{(\rho)}(x(t))\exp\qty[\int_0^t ds \Gamma(x(s),s)]
\end{align}
so that
\begin{align}
    \dot{\tilde{W}}_{t,1}^{(\rho)}(x(t)) = \Delta_0(x(t),t)\exp\qty[-\int_0^tds\Gamma(x(s),s)] \ .
\end{align}
Integrating this from $0$ to $t$ on both sides gives
\begin{align}
    W_{t,1}^{(\rho)}(x(t))-W_{0,1}^{(\rho)}(x_0) = \int_0^{t}du \Delta_0(x(u),u)\exp\qty[-\int_0^{u}ds \Gamma(x(s),s)] \ .
\end{align}
As in the case of $W_0$, the general solution is expressed in terms of the map $\Phi_t$ determining the characteristic curves as
\begin{align}
    W_{t,1}^{(\rho)}(x) = W_{0,1}^{(\rho)}(\Phi_t^{-1}(x)) + \int_0^{t}du \Delta_0(\Phi_u\circ\Phi_t^{-1}(x),u)\exp\qty[-\int_0^{u}ds \Gamma(\Phi_s\circ\Phi_t^{-1}(x),s)] \ . \label{eq:solution_for_W1}
\end{align}

\subsection{Solution for reverse process}
In this subsection, we construct a solution for reverse process. In the following discussion, we denote the Wigner function for reverse process by $\bar{W}^{(\rho)}$. The backward equation is given by Eq. \eqref{eq:1st order of Wigner tran of inverse Lindblad eq}.
Rearranging the backward drift $\bar{f}^{\mu}$ in Eq. \eqref{eq:1st order of Wigner tran of inverse Lindblad eq} order by order in $\hbar$, we obtain
\begin{align}
    \bar{f}^{\mu} = -\omega^{\mu\nu}\qty[\pdv{H_t}{x^\nu} + K_{\nu}(x,t)] + \hbar\qty[\frac{1}{W^{(\gamma)}_t}\pdv{}{x^\nu}\qty(D^{\mu\nu}(x,t)W^{(\gamma)}_t)+\frac{1}{2}\omega^{\mu\nu}J_{\nu}(x,t)] + \cdots \ .
\end{align}
As for the term corresponding to the score, since the Wigner function of the forward process can be expanded in powers of $\hbar$, we have
\begin{align}
    \frac{1}{W_t^{(\gamma)}}\pdv{}{x^{\nu}}\qty(D^{\mu\nu}(x,t)W_t^{(\gamma)}) = \frac{1}{W_{t,0}^{(\gamma)}}\pdv{}{
    x^{\nu}}\qty(D^{\mu\nu}(x,t)W_{t,0}^{(\gamma)}) + \order{\hbar} \ .
\end{align}
That is, the drift in the reverse process is
\begin{align}
    \bar{f}^{\mu}(x,t) = -\omega^{\mu\nu}\qty[\pdv{H_t}{x^\nu} + K_{\nu}(x,t)] + \hbar\qty[\frac{1}{W_{t,0}^{(\gamma)}(x)}\pdv{x^\nu}(D^{\mu\nu}(x,t)W_{t,0}^{(\gamma)})+\frac{1}{2}\omega^{\mu\nu}J_{\nu}(x,t)] + \order{\hbar^2} \ .
\end{align}
Thus, assuming that the Wigner function of the reverse process admits the expansion
\begin{align}
    \bar{W}_t^{(\rho)}(x) = \bar{W}^{(\rho)}_{t,0} + \hbar \bar{W}^{(\rho)}_{t,1} + \cdots
\end{align}
and organizing Eq. \eqref{eq:1st order of Wigner tran of inverse Lindblad eq} order by order in $\hbar$, we obtain
\begin{align}
    -\pdv{\bar{W}^{(\rho)}_{t,0}}{t} &= \pdv{x^\mu}\qty[\omega^{\mu\nu}\qty(\pdv{H_t}{x^\nu} + K_{\nu}(x,t))\bar{W}^{(\rho)}_{t,0}] \ , \\
    -\pdv{\bar{W}^{(\rho)}_{t,1}}{t} &= \pdv{x^\mu}\qty[\omega^{\mu\nu}\qty(\pdv{H_t}{x^\nu} + K_{\nu}(x,t))\bar{W}^{(\rho)}_{t,1}] \nonumber \\
    &\quad -\pdv{x^\mu} \qty[\qty(\frac{1}{W_{t,0}^{(\gamma)}}\pdv{x^\nu}(D^{\mu\nu}(x,t)W_{t,0}^{(\gamma)})+\frac{1}{2}\omega^{\mu\nu}J_{\nu}(x,t))\bar{W}^{(\rho)}_{t,0}] \nonumber \\
    &\quad + \frac{1}{2}\ppdiff{}{x^\mu}{x^\nu}\qty(D^{\mu\nu}(x,t)\bar{W}^{(\rho)}_{t,0}) \ .
\end{align}
Now suppose that $t$ satisfies $0\leq t \leq T$. In other words, let the final time of the forward process be $T$. Define $\bar{t}\coloneqq T-t$. Then, $t:0\to T$
corresponds to $\bar{t}:T\to 0$. Also, $\pdv{\bar{t}} = -\pdv{t}$.
Furthermore, define $\bar{x} = (\bar{Q}, \bar{P}) \coloneqq (Q, -P)$, and set
\begin{align}
    \bar{A}(\bar{Q},\bar{P},\bar{t}) &= -A(\bar{Q},-\bar{P},T-\bar{t}) \ ,\\
    \bar{B}(\bar{Q},\bar{P},\bar{t}) &= -B(\bar{Q},-\bar{P},T-\bar{t}) \ ,\\
    \bar{\Gamma}(\bar{Q},\bar{P},\bar{t}) &= -\Gamma(\bar{Q},-\bar{P},T-\bar{t}) \ .
\end{align}
Then $\pdv{Q} = \pdv{\bar{Q}}$ and $\pdv{P} = -\pdv{\bar{P}}$. Using these, we obtain
\begin{align}
    &\pdv{\bar{t}}\bar{W}^{(\rho)}_{\bar{t},0}(\bar{x}) +\bar{A}(\bar{x},\bar{t})\pdv{\bar{Q}}\bar{W}^{(\rho)}_{\bar{t},0}(\bar{x}) -\bar{B}(\bar{x},\bar{t})\pdv{\bar{P}}\bar{W}^{(\rho)}_{\bar{t},0}(\bar{x}) + \bar{\Gamma}(\bar{x},\bar{t})\bar{W}^{(\rho)}_{\bar{t},0}(\bar{x}) = 0 \ ,\label{eq:backward_0th_order_of_hbar_2}\\
    &\pdv{\bar{t}}\bar{W}^{(\rho)}_{\bar{t},1}(\bar{x}) +\bar{A}(\bar{x},\bar{t})\pdv{\bar{Q}}\bar{W}^{(\rho)}_{\bar{t},1}(\bar{x}) - \bar{B}(\bar{x},\bar{t})\pdv{\bar{P}}\bar{W}^{(\rho)}_{\bar{t},1}(\bar{x}) + \bar{\Gamma}(\bar{x},\bar{t})\bar{W}^{(\rho)}_{\bar{t},1}(\bar{x}) = \bar{\Delta}_0(\bar{x},\bar{t}) \ .
    \label{eq:backward_1th_order_of_hbar_2}
\end{align}
Note that we have rewritten $\bar{W}^{(\rho)}_{T-\bar{t}}(\bar{Q},-\bar{P})$ simply as $\bar{W}^{(\rho)}_{\bar{t}}(\bar{x})$. We also have
\begin{align}
    \bar{\Delta}_0(\bar{x},\bar{t}) &\coloneqq \left.-\frac{1}{2}\qty[\pdv{x^\mu}\qty{\omega^{\mu\nu}J_{\nu}(x,t)\bar{W}^{(\rho)}_{t,0}} - \ppdiff{}{x^\mu}{x^\nu}(D^{\mu\nu}(x,t)\bar{W}^{(\rho)}_{t,0})]\right|_{Q=-\bar{Q},P=-\bar{P},t=T-\bar{t}} \nonumber \\
    &\quad \left.-\pdv{x^\mu}\qty[\qty(\frac{1}{W_{t,0}^{(\gamma)}}\pdv{x^\nu}\qty(D^{\mu\nu}(x,t)W_{t,0}^{(\gamma)}))\bar{W}^{(\rho)}_{t,0}]\right|_{Q=\bar{Q},P=-\bar{P},t=T-\bar{t}} \ .
\end{align}
Here, unlike $A$, $B$ and $\Gamma$, we do not have $\bar{\Delta}_0 = \left.-\Delta_0\right|_{Q=\bar{Q},P=-\bar{P},t=T-\bar{t}}$. On the other hand, when $\bar{W}^{(\rho)}_0 = W_0^{(\gamma)}$, we have $\bar{\Delta}_0 = \left.-\Delta_0\right|_{Q=\bar{Q},P=-\bar{P},t=T-\bar{t}}$.
Denoting the total derivative with respect to $\bar{t}$ by a white dot, such as  $\whitedot{f}$, we now consider solving Eqs. \eqref{eq:backward_0th_order_of_hbar_2} and \eqref{eq:backward_1th_order_of_hbar_2}. As in the forward-process case, these can be solved by the method of characteristics. The characteristic equations are
\begin{align}
    \whitedot{\bar{Q}} = \bar{A}(\bar{x}(\bar{t}),\bar{t}), \quad \whitedot{\bar{P}} = -\bar{B}(\bar{x}(\bar{t}),\bar{t}) \ .
\end{align}
Solving these yields the map
\begin{align}
    \Psi_{\bar{t}}: (\bar{x}(\bar{t}=0)) \to \bar{x}(\bar{t}) \ .
\label{Psi}
\end{align}
The solution of the equation is then constructed in exactly the same manner as in the forward-process case, using $\Psi_{\bar{t}}$ instead of $\Phi_{\bar{t}}$.

\section{Conclusion and outlook}
In this paper, we have shown that
applying the classical
approximation, which reduces a
Lindblad equation
to the Fokker–Planck equation for the Wigner function, to the Lindblad equation associated with the Petz map yields another equation for the Wigner function that coincides with that obtained from the Fokker–Planck equation via Bayes’ rule.
This establishes a direct correspondence between the Petz map and Bayes’ rule, thereby unifying quantum reversibility with classical reverse diffusion.
We also construct WKB solutions for the
semi-classical Lindblad and reverse
semi-classical Lindblad equations.

The correspondence between the Petz map and reversal diffusion
equations has been investigated in
specific set-ups where
certain classicalization is realized.
In fact, it was shown in
Ref. \cite{Hu:2025lyk} that when $N=1$, $\hat{H}=0$ and $\hat{L}=\hat{p}$,
the Lindblad and reversal Lindblad equations are closed for
decoherent states
for which these two equations
exactly correspond
to the diffusion and reversal diffusion equations, respectively.
It was also shown in Ref. \cite{Liu:2025spz} that when
the measurement-based Lindblad dynamics is applied to
an $n$-qubit separable state expressed as a mixture of tensor
product of spin coherent states, the Petz map reduces
to a classical reversal diffusion equation in the large-spin (classical) limit.
On the other hand,
compared to the above studies,
the correspondence found in this paper
holds in the semiclassical limit, namely up to the next-leading order in the
$\hbar$ expansion, for
generic systems where the number
of degrees of freedom, the Wigner function, Hamiltonian and Lindblad operators are all arbitrary.

A possible
future direction is extending the correspondence found in this paper to mesoscopic systems
and exploring its implications for constructing quantum analogues of diffusion-based generative models \cite{Parigi:2023djo, Cacioppo:2023xdq, Chen:2024zye, Kolle:2024mhw, Zhang:2023jxy, Zhu:2024vep, tang2025quadim, Kwun:2024edh, liu2025beyond, Huang:2025wrk, ZHANG2026107981, Cui:2025sxu, Liu:2025spz, Hu:2025lyk}.

\section*{Acknowledgements}
We would like to thank Tatsuro Yuge and Kazuki Kobayashi for discussions.
 A.T. was supported in
 part by JSPS KAKENHI Grant Numbers 21K03532 and 25K07319.

\appendix
\renewcommand{\theequation}{\Alph{section}.\arabic{equation}}

\section{$\hbar$ expansion}
In Appendix A, $[\;]_W$ denotes the Wigner transform.
For instance,
\begin{align}
    [\hat{O}]_W &= O(\bm{Q},\bm{P}) \ , \nonumber\\
    [\hat{\mathcal{O}}_1\hat{\mathcal{O}}_2]_W
    &= \mathcal{O}_1(\bm{Q},\bm{P}) \star \mathcal{O}_2(\bm{Q},\bm{P}) \ .
    \nonumber
\end{align}

\subsection{$\hbar$ expansion of Lindblad equation}
\label{appendix:hbar expansion of Lindblad eq}
In this appendix,
we perform the Wigner transformation of the Lindblad equation \eqref{eq:lindblad} and
expand the Wigner transform up to $\mathcal{O}(\hbar)$ using Eqs. \eqref{Moyal product}
and \eqref{hbarexpansion}.

First, the left-hand side of Eq. \eqref{eq:lindblad} transforms as
\begin{align}
    \left[\pdv{\oprho_{t}}{t}\right]_{W} = \pdv{W_{t}^{(\rho)}}{t} \ .
\end{align}
Second, we perform the Wigner transformation of the third term on the
right-hand side of Eq. \eqref{eq:lindblad}. By using an $\hbar$ expansion
\begin{align}
    &\hbar\left[\hat{L}_{\alpha,t}^{\dagger}\hat{L}_{\alpha,t}\right]_{W} = \ell_{\alpha,t}^{*}\star\ell_{\alpha,t} \nonumber\\
    &=\abs{\ell_{\alpha,t}}^{2} + \frac{i\hbar}{2}\left\{\ell^{*}_{\alpha,t},\ell_{\alpha,t}\right\}_{p} - \frac{\hbar^2}{8}\sum_{i}\left[\left\{\pdiff{\ell_{\alpha,t}^{*}}{Q_{i}},\pdiff{\ell_{\alpha,t}}{P_{i}}\right\}_{p} - \left\{\pdiff{\ell_{\alpha,t}^{*}}{P_{i}},\pdiff{\ell_{\alpha,t}}{Q_{i}}\right\}_{p}\right] + \mathcal{O}(\hbar^{3}) \ ,
\end{align}
we obtain
\begin{align}
&\hbar\left[\hat{L}_{\alpha,t}^{\dagger}\hat{L}_{\alpha,t}\oprho_{t}\right]_{W} = \ell_{\alpha,t}^{*}\star\ell_{\alpha,t}\star W_{t}^{(\rho)} \nonumber\\
    &\begin{aligned}
        &=\abs{\ell_{\alpha,t}}^2 W_{t}^{(\rho)} + \frac{i\hbar}{2} \qty{\abs{\ell_{\alpha,t}}^2, W_{t}^{(\rho)}}_p
        - \frac{\hbar^2}{8} \sum_i \qty(
        \qty{\frac{\partial \abs{\ell_{\alpha,t}}^2}{\partial Q_i}, \frac{\partial W_{t}^{(\rho)}}{\partial P_i}}_{p}
        - \qty{\frac{\partial \abs{\ell_{\alpha,t}}^2}{\partial P_i}, \frac{\partial W_{t}^{(\rho)}}{\partial Q_i}}_{p}) \\
        &\quad + \frac{i\hbar}{2} \qty( \qty{\ell_{\alpha,t}^*, \ell_{\alpha,t}}_p W_{t}^{(\rho)} + \frac{i\hbar}{2} \qty{ \qty{\ell_{\alpha,t}^*, \ell_{\alpha,t}}_p, W_{t}^{(\rho)} }_p ) \\
        &\quad - \frac{\hbar^2}{8} \sum_i \qty(
        \qty{\frac{\partial \ell_{\alpha,t}^*}{\partial Q_i}, \frac{\partial \ell_{\alpha,t}}{\partial P_i}}_{p}
        - \qty{\frac{\partial \ell_{\alpha,t}^*}{\partial P_i}, \frac{\partial \ell_{\alpha,t}}{\partial Q_i}}_{p} ) W_{t}^{(\rho)}
        + \order{\hbar^3} \ .
    \end{aligned}  \label{temp1}
\end{align}
A similar calculation yields
\begin{align}
&\hbar\left[\oprho_{t}\hat{L}_{\alpha,t}^{\dagger}\hat{L}_{\alpha,t}\right]_{W} = W_{t}^{(\rho)}\star\ell_{\alpha,t}^{*}\star\ell_{\alpha,t} \nonumber\\
    &\begin{aligned}
    &=W_{t}^{(\rho)}\abs{\ell_{\alpha,t}}^2  + \frac{i\hbar}{2} \qty{W_{t}^{(\rho)}, \abs{\ell_{\alpha,t}}^2}_p
        - \frac{\hbar^2}{8} \sum_i \qty(
        \qty{\frac{\partial W_{t}^{(\rho)}}{\partial Q_i}, \frac{\partial \abs{\ell_{\alpha,t}}^2}{\partial P_i}}_{p} - \qty{\frac{\partial W_{t}^{(\rho)}}{\partial P_i}, \frac{\partial \abs{\ell_{\alpha,t}}^2}{\partial Q_i}}_{p} ) \\
        &\quad + \frac{i\hbar}{2} \qty(W_{t}^{(\rho)}\qty{\ell_{\alpha,t}^*, \ell_{\alpha,t}}_p + \frac{i\hbar}{2} \qty{W_{t}^{(\rho)} , \qty{\ell_{\alpha,t}^*, \ell_{\alpha,t}}_p}_p ) \\
        &\quad - W_{t}^{(\rho)}\frac{\hbar^2}{8} \sum_i \qty(
        \qty{\frac{\partial \ell_{\alpha,t}^*}{\partial Q_i}, \frac{\partial \ell_{\alpha,t}}{\partial P_i}}_{p}
        - \qty{\frac{\partial \ell_{\alpha,t}^*}{\partial P_i}, \frac{\partial \ell_{\alpha,t}}{\partial Q_i}}_{p} )
        + \order{\hbar^3} \ .
    \end{aligned} \label{temp2}
\end{align}
Combining Eqs. \eqref{temp1} and \eqref{temp2} leads to
\begin{align}
    &\hbar\left[-\frac{1}{2}\left[\hat{L}^{\dagger}_{\alpha,t}\hat{L}_{\alpha,t},\oprho_{t}\right]_{+}\right]_{W} = -\frac{1}{2\hbar}\qty(\ell_{\alpha,t}^{*}\star\ell_{\alpha,t}\star W_{t}^{(\rho)}+W_{t}^{(\rho)}\star\ell_{\alpha,t}^{*}\star\ell_{\alpha,t}) \nonumber\\
    &=-\frac{1}{\hbar}\left.\ell_{\alpha,t}^{*}\star\ell_{\alpha,t}\right|_{\leq\hbar^2}W_{t}^{(\rho)} + \frac{\hbar}{8} \sum_i \qty(
        \qty{\frac{\partial W_{t}^{(\rho)}}{\partial Q_i}, \frac{\partial \abs{\ell_{\alpha,t}}^2}{\partial P_i}}_{p} - \qty{\frac{\partial W_{t}^{(\rho)}}{\partial P_i}, \frac{\partial \abs{\ell_{\alpha,t}}^2}{\partial Q_i}}_{p} ) + \order{\hbar^2} \ .
\end{align}
Here, we have defined $\left.\ell_{\alpha,t}^{*}\star\ell_{\alpha,t}\right|_{\leq\hbar^2}$ by
\begin{align}
    \left.\ell_{\alpha,t}^{*}\star\ell_{\alpha,t}\right|_{\leq\hbar^2} =\abs{\ell_{\alpha,t}}^2 + \frac{i\hbar}{2}\qty{\ell^{*}_{\alpha,t},\ell_{\alpha,t}}_{p} -\frac{\hbar^2}{8} \sum_i \qty(
        \qty{\frac{\partial \ell_{\alpha,t}^*}{\partial Q_i}, \frac{\partial \ell_{\alpha,t}}{\partial P_i}}_{p}
        - \qty{\frac{\partial \ell_{\alpha,t}^*}{\partial P_i}, \frac{\partial \ell_{\alpha,t}}{\partial Q_i}}_{p} ) \ .\label{ll star product by hbar2}
\end{align}
Third, the Wigner transformation of the second term on the right-hand side of
Eq. \eqref{eq:lindblad} reads
\begin{align}
&\hbar\left[\hat{L}_{\alpha,t}\oprho_{t}\hat{L}^{\dagger}_{\alpha,t}\right]_{W}
\nonumber\\
    &=\ell_{\alpha,t}\star\left[W_{t}^{(\rho)} \ell_{\alpha,t}^* + \frac{i\hbar}{2} \qty{W_{t}^{(\rho)}, \ell_{\alpha,t}^*}_p  - \frac{\hbar^2}{8} \sum_i \qty(
    \qty{\frac{\partial W_{t}^{(\rho)}}{\partial Q_i}, \frac{\partial \ell_{\alpha,t}^*}{\partial P_i}}_p
    - \qty{\frac{\partial W_{t}^{(\rho)}}{\partial P_i}, \frac{\partial \ell_{\alpha,t}^*}{\partial Q_i}}_p ) \right] + \order{\hbar^3} \nonumber \\
    &\begin{aligned}
        &=\abs{\ell_{\alpha,t}}^2 W_{t}^{(\rho)} + \frac{i\hbar}{2} \qty{\ell_{\alpha,t}, W_{t}^{(\rho)} \ell_{\alpha,t}^*}_p - \frac{\hbar^2}{8} \sum_i \qty(
        \qty{\frac{\partial \ell_{\alpha,t}}{\partial Q_i}, \frac{\partial (W_{t}^{(\rho)} \ell_{\alpha,t}^*)}{\partial P_i}}_p
        - \qty{\frac{\partial \ell_{\alpha,t}}{\partial P_i}, \frac{\partial (W_{t}^{(\rho)} \ell_{\alpha,t}^*)}{\partial Q_i}}_p ) \\
        &\quad + \frac{i\hbar}{2} \qty[
        \ell_{\alpha,t} \qty{W_{t}^{(\rho)}, \ell_{\alpha,t}^*}_p + \frac{i\hbar}{2} \qty{\ell_{\alpha,t}, \qty{W_{t}^{(\rho)}, \ell_{\alpha,t}^*}_p}_p] \\
        &\quad - \frac{\hbar^2}{8} \sum_i \ell_{\alpha,t} \qty(
        \qty{\frac{\partial W_{t}^{(\rho)}}{\partial Q_i}, \frac{\partial \ell_{\alpha,t}^*}{\partial P_i}}_p
        - \qty{\frac{\partial W_{t}^{(\rho)}}{\partial P_i}, \frac{\partial \ell_{\alpha,t}^*}{\partial Q_i}}_p ) + \order{\hbar^3} \ . \label{eq:temp60}
\end{aligned}
\end{align}
In Eq. \eqref{eq:temp60}, the terms of first order in $\hbar$  are collected as
\begin{align}
    &\frac{i\hbar}{2} \qty{\ell_{\alpha,t}, W_{t}^{(\rho)} \ell_{\alpha,t}^*}_p
+ \frac{i\hbar}{2} \ell_{\alpha,t} \qty{W_{t}^{(\rho)}, \ell_{\alpha,t}^*}_p  \nonumber\\
&= \frac{i\hbar}{2} \qty(
  \qty{\ell_{\alpha,t}, W_{t}^{(\rho)}}_p \ell_{\alpha,t}^*
  + W_{t}^{(\rho)} \qty{\ell_{\alpha,t}, \ell_{\alpha,t}^*}_p
  + \ell_{\alpha,t} \qty{W_{t}^{(\rho)}, \ell_{\alpha,t}^*}_p ) \nonumber\\
&=- \frac{i\hbar}{2} \qty{\ell_{\alpha,t}, \ell_{\alpha,t}^*}_p W_{t}^{(\rho)}
- \hbar \sum_i \pdv{Q_i} \qty[
  \Im \qty( \ell_{\alpha,t} \pdv{\ell_{\alpha,t}^*}{P_i} ) W_{t}^{(\rho)}
] - \hbar \sum_i \pdv{P_i} \qty[
  \Im \qty( \pdv{\ell_{\alpha,t}}{Q_i} \ell_{\alpha,t}^* ) W_{t}^{(\rho)}
] \ .\label{eq:temp61}
\end{align}
Here, the Leibniz rule of the Poisson bracket has been used in the first equality.
The terms of second order in $\hbar$ are collected as
\begin{align}
&\begin{aligned}
    &- \frac{\hbar^2}{8} \sum_i \qty(
    \qty{\frac{\partial \ell_{\alpha,t}}{\partial Q_i}, \frac{\partial (W_{t}^{(\rho)} \ell_{\alpha,t}^*)}{\partial P_i}}_p
    - \qty{\frac{\partial \ell_{\alpha,t}}{\partial P_i}, \frac{\partial (W_{t}^{(\rho)} \ell_{\alpha,t}^*)}{\partial Q_i}}_p ) + \frac{i\hbar}{2} \frac{i\hbar}{2} \qty{\ell_{\alpha,t}, \qty{W_{t}^{(\rho)}, \ell_{\alpha,t}^*}_p}_p \\
    &\quad - \frac{\hbar^2}{8} \sum_i \ell_{\alpha,t} \qty(
    \qty{\frac{\partial W_{t}^{(\rho)}}{\partial Q_i}, \frac{\partial \ell_{\alpha,t}^*}{\partial P_i}}_p
    - \qty{\frac{\partial W_{t}^{(\rho)}}{\partial P_i}, \frac{\partial \ell_{\alpha,t}^*}{\partial Q_i}}_p )
\end{aligned}\nonumber\\
&\begin{aligned}
& = - \frac{\hbar^2}{8} \sum_i \left( \left\{ \frac{\partial \ell_{\alpha,t}}{\partial Q_i}, \frac{\partial \ell_{\alpha,t}^*}{\partial P_i} \right\}_p - \left\{ \frac{\partial \ell_{\alpha,t}}{\partial P_i}, \frac{\partial \ell_{\alpha,t}^*}{\partial Q_i} \right\}_p \right) W_{t}^{(\rho)} \\
    &\qquad - \frac{\hbar^2}{2} \sum_i \left( \Re \left\{ \ell_{\alpha,t}, \frac{\partial \ell_{\alpha,t}^*}{\partial P_i} \right\}_p \frac{\partial W_{t}^{(\rho)}}{\partial Q_i} + \mathrm{Re} \left\{ \frac{\partial \ell_{\alpha,t}^*}{\partial Q_i}, \ell_{\alpha,t} \right\}_p \frac{\partial W_{t}^{(\rho)}}{\partial P_i} \right) \\
    &\qquad - \frac{\hbar^2}{4} \sum_{i,j} \Biggl[ -\frac{\partial \ell_{\alpha,t}}{\partial Q_j} \frac{\partial \ell_{\alpha,t}^*}{\partial Q_i} \frac{\partial^2 W_{t}^{(\rho)}}{\partial P_i \partial P_j} + \left( \frac{\partial \ell_{\alpha,t}}{\partial Q_j} \frac{\partial \ell_{\alpha,t}^*}{\partial P_i} + \frac{\partial \ell_{\alpha,t}}{\partial P_i} \frac{\partial \ell_{\alpha,t}^*}{\partial Q_j} \right) \frac{\partial^2 W_{t}^{(\rho)}}{\partial Q_i \partial P_j}  - \frac{\partial \ell_{\alpha,t}}{\partial P_j} \frac{\partial \ell_{\alpha,t}^*}{\partial P_i} \frac{\partial^2 W_{t}^{(\rho)}}{\partial Q_i \partial Q_j} \Biggr] \\
    & \qquad- \frac{\hbar^2}{8} \sum_i \left[ \ell_{\alpha,t} \left( \left\{ \frac{\partial W_{t}^{(\rho)}}{\partial Q_i}, \frac{\partial \ell_{\alpha,t}^*}{\partial P_i} \right\}_p - \left\{ \frac{\partial W_{t}^{(\rho)}}{\partial P_i}, \frac{\partial \ell_{\alpha,t}^*}{\partial Q_i} \right\}_p \right) + \text{c.c.} \right] \ .
    \end{aligned}\label{eq:temp62}
\end{align}
By substituting Eqs. \eqref{eq:temp61} and \eqref{eq:temp62}
into Eq. \eqref{eq:temp60}, we obtain
\begin{align}
    &\hbar\left[\hat{L}_{\alpha,t}\oprho\hat{L}_{\alpha,t}^{\dagger}\right]_{W} \\
    &\begin{aligned}
        &=\left.\ell_{\alpha,t}^{*}\star\ell_{\alpha,t}\right|_{\leq\hbar^2}\\
        &\qquad- \hbar \sum_i \pdv{Q_i} \qty[
        \Im \qty( \ell_{\alpha,t} \pdv{\ell_{\alpha,t}^*}{P_i} ) W_{t}^{(\rho)}
        ] - \hbar \sum_i \pdv{P_i} \qty[
        \Im \qty( \pdv{\ell_{\alpha,t}}{Q_i} \ell_{\alpha,t}^* ) W_{t}^{(\rho)}
        ]\\
        &\qquad - \frac{\hbar^2}{2} \sum_i \left( \Re \left\{ \ell_{\alpha,t}, \frac{\partial \ell_{\alpha,t}^*}{\partial P_i} \right\}_p \frac{\partial W_{t}^{(\rho)}}{\partial Q_i} + \mathrm{Re} \left\{ \frac{\partial \ell_{\alpha,t}^*}{\partial Q_i}, \ell_{\alpha,t} \right\}_p \frac{\partial W_{t}^{(\rho)}}{\partial P_i} \right) \\
        &\qquad - \frac{\hbar^2}{4} \sum_{i,j} \Biggl[ -\frac{\partial \ell_{\alpha,t}}{\partial Q_j} \frac{\partial \ell_{\alpha,t}^*}{\partial Q_i} \frac{\partial^2 W_{t}^{(\rho)}}{\partial P_i \partial P_j} + \left( \frac{\partial \ell_{\alpha,t}}{\partial Q_j} \frac{\partial \ell_{\alpha,t}^*}{\partial P_i} + \frac{\partial \ell_{\alpha,t}}{\partial P_i} \frac{\partial \ell_{\alpha,t}^*}{\partial Q_j} \right) \frac{\partial^2 W_{t}^{(\rho)}}{\partial Q_i \partial P_j}  - \frac{\partial \ell_{\alpha,t}}{\partial P_j} \frac{\partial \ell_{\alpha,t}^*}{\partial P_i} \frac{\partial^2 W_{t}^{(\rho)}}{\partial Q_i \partial Q_j} \Biggr] \\
        & \qquad- \frac{\hbar^2}{8} \sum_i \left[ \ell_{\alpha,t} \left( \left\{ \frac{\partial W_{t}^{(\rho)}}{\partial Q_i}, \frac{\partial \ell_{\alpha,t}^*}{\partial P_i} \right\}_p - \left\{ \frac{\partial W_{t}^{(\rho)}}{\partial P_i}, \frac{\partial \ell_{\alpha,t}^*}{\partial Q_i} \right\}_p \right) + \text{c.c.} \right] \ .
    \end{aligned}
\end{align}
Therefore, we can expand the right-hand side of Eq. \eqref{eq:lindblad} as
\begin{align}
    &\hbar\left[\sum_{\alpha}\left(\hat{L}_{\alpha,t}\oprho_{t}\hat{L}_{\alpha,t}^{\dagger}-\frac{1}{2}\left[\hat{L}_{\alpha,t}^{\dagger}\hat{L}_{\alpha,t},\oprho_{t}\right]_{+}\right)\right]_{W}\nonumber\\
    &\begin{aligned}
    = \sum_\alpha &\left[  - \sum_i \frac{\partial}{\partial Q_i} \left[ \mathrm{Im} \left( \ell_{\alpha,t} \frac{\partial \ell_{\alpha,t}^*}{\partial P_i} \right) W_{t}^{(\rho)} \right] - \sum_i \frac{\partial}{\partial P_i} \left[ \mathrm{Im} \left( \frac{\partial \ell_{\alpha,t}}{\partial Q_i} \ell_{\alpha,t}^* \right) W_{t}^{(\rho)} \right]\right. \\
    & - \frac{\hbar}{2} \sum_i \left( \left(\mathrm{Re}\left\{ \ell_{\alpha,t}, \frac{\partial \ell_{\alpha,t}^*}{\partial P_i} \right\}_p\frac{\partial W_{t}^{(\rho)}}{\partial Q_i} + \mathrm{Re}\left\{ \frac{\partial \ell_{\alpha,t}^*}{\partial Q_i}, \ell_{\alpha,t} \right\}_p \frac{\partial W_{t}^{(\rho)}}{\partial P_i} \right) \right.\\
    & + \frac{1}{2} \sum_{i,j} \Biggl\{ - \frac{\partial \ell_{\alpha,t}}{\partial Q_j} \frac{\partial \ell_{\alpha,t}^*}{\partial Q_i} \frac{\partial^2 W_{t}^{(\rho)}}{\partial P_i \partial P_j} + \left( \frac{\partial \ell_{\alpha,t}}{\partial Q_j} \frac{\partial \ell_{\alpha,t}^*}{\partial P_i} + \frac{\partial \ell_{\alpha,t}}{\partial P_i} \frac{\partial \ell_{\alpha,t}^*}{\partial Q_j} \right) \frac{\partial^2 W_{t}^{(\rho)}}{\partial Q_i \partial P_j} \\
    & \hphantom{+ \frac{1}{2} \sum_{i,j} \Biggl\{} - \frac{\partial \ell_{\alpha,t}}{\partial P_j} \frac{\partial \ell_{\alpha,t}^*}{\partial P_i} \frac{\partial^2 W_{t}^{(\rho)}}{\partial Q_i \partial Q_j} \Biggr\} \\
    & + \frac{1}{4} \sum_i \left\{ \ell_{\alpha,t} \left( \left\{ \frac{\partial W_{t}^{(\rho)}}{\partial Q_i}, \frac{\partial \ell_{\alpha,t}^*}{\partial P_i} \right\}_p - \left\{ \frac{\partial W_{t}^{(\rho)}}{\partial P_i}, \frac{\partial \ell_{\alpha,t}^*}{\partial Q_i} \right\}_p \right) + \text{c.c.} \right\} \\
    & \left.\left.- \frac{1}{4} \sum_i \left( \left\{ \frac{\partial W_{t}^{(\rho)}}{\partial Q_i}, \frac{\partial |\ell_{\alpha,t}|^2}{\partial P_i} \right\}_p - \left\{ \frac{\partial W_{t}^{(\rho)}}{\partial P_i}, \frac{\partial |\ell_{\alpha,t}|^2}{\partial Q_i} \right\}_p \right) \right)\right] + \order{\hbar^2} \ .
\end{aligned}\label{eq:temp70}
\end{align}
In this equation, we can simplify the terms of the first order in $\hbar$ further as
\begin{align}
    &\begin{aligned}
    &  - \frac{\hbar}{2} \sum_i \left[ \left(\sum_\alpha \mathrm{Re}\left\{ \ell_{\alpha,t}, \frac{\partial \ell_{\alpha,t}^*}{\partial P_i} \right\}_p\right)\frac{\partial W_{t}^{(\rho)}}{\partial Q_i} + \left(\sum_\alpha \mathrm{Re} \left\{ \frac{\partial \ell_{\alpha,t}^*}{\partial Q_i}, \ell_{\alpha,t} \right\}_p\right) \frac{\partial W_{t}^{(\rho)}}{\partial P_i} \right]\\
    &\qquad + \frac{\hbar}{2} \sum_{\mu,\nu} D_R^{\mu\nu}(\bm{x},t) \frac{\partial^2 W_{t}^{(\rho)}}{\partial x^{\mu} \partial x^{\nu}} \ .
    \end{aligned}\label{eq:temp80}
\end{align}
Here, we have introduced a $2N\times 2N$ Hermitian matrix
\begin{align}
    D^{\mu\nu}_c(\bm{x},t) = \sum_\alpha
    \renewcommand{\arraystretch}{2.5}
    \begin{pmatrix}
        \displaystyle \dfrac{\partial \ell_{\alpha,t}}{\partial P_\mu} \dfrac{\partial \ell_{\alpha,t}^*}{\partial P_\nu} & -\displaystyle \dfrac{\partial \ell_{\alpha,t}}{\partial P_\mu} \dfrac{\partial \ell_{\alpha,t}^*}{\partial Q_{\nu-N}} \\
        - \displaystyle \dfrac{\partial \ell_{\alpha,t}^*}{\partial P_\nu} \dfrac{\partial \ell_{\alpha,t}}{\partial Q_{\mu-N}} & \displaystyle \dfrac{\partial \ell_{\alpha,t}}{\partial Q_{\mu-N}} \dfrac{\partial \ell_{\alpha,t}^*}{\partial Q_{\nu-N}}
    \end{pmatrix} \ ,
\end{align}
and defined its real part by $D_R^{\mu\nu}= \Re(D_c^{\mu\nu}) = (D_c^{\mu\nu} +D_c^{\nu\mu})/2$.
By using Leibniz's rule, we rewrite Eq. \eqref{eq:temp80}
in the form of the total derivative as follows:
\begin{align}
    & - \frac{\hbar}{2} \sum_i\left[ \left( \sum_\alpha \mathrm{Re} \left\{ \ell_{\alpha,t}, \frac{\partial \ell_{\alpha,t}^*}{\partial P_i} \right\}_p \right) \frac{\partial W_{t}^{(\rho)}}{\partial Q_i} + \left( \sum_\alpha \mathrm{Re} \left\{ \frac{\partial \ell_{\alpha,t}^*}{\partial Q_i}, \ell_{\alpha,t} \right\}_p \right) \frac{\partial W_{t}^{(\rho)}}{\partial P_i} \right] \nonumber\\
    &\begin{aligned}
    &= - \frac{\hbar}{2} \sum_i \left[ \frac{\partial}{\partial Q_i} \left( \sum_\alpha \mathrm{Re} \left\{ \ell_{\alpha,t}, \frac{\partial \ell_{\alpha,t}^*}{\partial P_i} \right\}_p W_{t}^{(\rho)} \right) + \frac{\partial}{\partial P_i} \left( \sum_\alpha \mathrm{Re} \left\{ \frac{\partial \ell_{\alpha,t}^*}{\partial Q_i}, \ell_{\alpha,t} \right\}_p W_{t}^{(\rho)} \right) \right] \\
    & \qquad + \frac{\hbar}{2} W_{t}^{(\rho)} \sum_\alpha \sum_i \left[ \frac{\partial}{\partial Q_i} \left( \mathrm{Re} \left\{ \ell_{\alpha,t}, \frac{\partial \ell_{\alpha,t}^*}{\partial P_i} \right\}_p \right) + \frac{\partial}{\partial P_i} \left( \mathrm{Re} \left\{ \frac{\partial \ell_{\alpha,t}^*}{\partial Q_i}, \ell_{\alpha,t} \right\}_p \right) \right] \nonumber\\
    \end{aligned}\\
    &\begin{aligned}
        &= - \frac{\hbar}{2} \sum_i \left[ \frac{\partial}{\partial Q_i} \left( \sum_\alpha \mathrm{Re} \left\{ \ell_{\alpha,t}, \frac{\partial \ell_{\alpha,t}^*}{\partial P_i} \right\}_p W_{t}^{(\rho)} \right) + \frac{\partial}{\partial P_i} \left( \sum_\alpha \mathrm{Re} \left\{ \frac{\partial \ell_{\alpha,t}^*}{\partial Q_i}, \ell_{\alpha,t} \right\}_p  W_{t}^{(\rho)} \right) \right] \\
        & \qquad + \frac{\hbar}{2} W_{t}^{(\rho)} \sum_\alpha \sum_{i,j} \Biggl[ \frac{\partial^2 \ell_{\alpha,t}}{\partial Q_i \partial Q_j} \frac{\partial^2 \ell_{\alpha,t}^*}{\partial P_i \partial P_j} + \frac{\partial^2 \ell_{\alpha,t}^*}{\partial Q_i \partial Q_j} \frac{\partial^2 \ell_{\alpha,t}}{\partial P_i \partial P_j} - 2 \frac{\partial^2 \ell_{\alpha,t}}{\partial Q_i \partial P_j} \frac{\partial^2 \ell_{\alpha,t}^*}{\partial P_i \partial Q_j} \Biggr] \ ,
    \end{aligned}
\end{align}
and
\begin{align}
    \sum_{\mu,\nu}&D_{R}^{\mu\nu}\pdv{W_{t}^{(\rho)}}{x^{\mu}}{x^{\nu}}\nonumber\\
    &= \sum_{\mu,\nu} \frac{\partial^2}{\partial x^{\mu} \partial x^{\nu}} (D_{R}^{\mu\nu} W_{t}^{(\rho)}) - 2 \sum_{\mu,\nu} \frac{\partial D_{R}^{\mu\nu}}{\partial x^{\nu}} \frac{\partial W_{t}^{(\rho)}}{\partial x^{\mu}} - \sum_{\mu,\nu} \frac{\partial^2 D_{R}^{\mu\nu}}{\partial x^{\mu} \partial x^{\nu}} W_{t}^{(\rho)} \nonumber\\
    &\begin{aligned}
        &= \sum_{\mu,\nu} \frac{\partial^2}{\partial x^{\mu} \partial x^{\nu}} (D_{R}^{\mu\nu} W_{t}^{(\rho)}) - 2 \sum_{\mu,\nu} \frac{\partial}{\partial x^{\mu}} \left( \frac{\partial D_{R}^{\mu\nu}}{\partial x^{\nu}} W_{t}^{(\rho)} \right) + \sum_{\mu,\nu} \frac{\partial^2 D_{R}^{\mu\nu}}{\partial x^{\mu} \partial x^{\nu}} W_{t}^{(\rho)}
    \end{aligned}  \label{temp3}
\end{align}
Here, we have used $D_{R}^{\mu\nu} = D_{R}^{\nu\mu}$.
We calculate each term in Eq. \eqref{temp3}.
The third term reads
\begin{align}
    \sum_{\mu,\nu} &\frac{\partial^2 D_{R}^{\mu\nu}}{\partial x^{\mu} \partial x^{\nu}} = \sum_{\mu,\nu} \frac{\partial^2 D_c^{\mu\nu}}{\partial x^{\mu} \partial x^{\nu}} \nonumber\\
    &= \begin{aligned}[t]
        & \sum_{i,j} \frac{\partial^2}{\partial Q_i \partial Q_j} \left( \sum_\alpha \frac{\partial \ell_{\alpha,t}}{\partial P_i} \frac{\partial \ell_{\alpha,t}^*}{\partial P_j} \right) + \sum_{i,j} \frac{\partial^2}{\partial Q_i \partial P_j} \left[ 2 \mathrm{Re} \left( -\sum_\alpha \frac{\partial \ell_{\alpha,t}}{\partial P_i} \frac{\partial \ell_{\alpha,t}^*}{\partial Q_j} \right) \right] \nonumber\\
        & + \sum_{i,j} \frac{\partial^2}{\partial P_i \partial P_j} \left( \sum_\alpha\frac{\partial \ell_{\alpha,t}}{\partial Q_i} \frac{\partial \ell_{\alpha,t}^*}{\partial Q_j} \right)
    \end{aligned}\\
    &\begin{aligned}[t]
        &= \sum_{i,j} \sum_\alpha \Biggl[ 2 \frac{\partial^2 \ell_{\alpha,t}}{\partial Q_j \partial P_i} \frac{\partial^2 \ell_{\alpha,t}^*}{\partial Q_i \partial P_j} - \left( \frac{\partial^2 \ell_{\alpha,t}}{\partial P_i \partial P_j} \frac{\partial^2 \ell_{\alpha,t}^*}{\partial Q_i \partial Q_j} + \frac{\partial^2 \ell_{\alpha,t}^*}{\partial P_i \partial P_j} \frac{\partial^2 \ell_{\alpha,t}}{\partial Q_i \partial Q_j} \right) \Biggr] \ .
    \end{aligned}
\end{align}
The second term reads
\begin{align}
    &\sum_{\mu,\nu}\pdv{}{x^{\mu}}\left(\pdv{D_{R}^{\mu\nu}}{x^{\nu}}W_{t}^{(\rho)}\right) \nonumber\\
    &\begin{aligned}[t]
        &= \sum_{i,j} \frac{\partial}{\partial Q_i} \left[ \left\{ \frac{\partial}{\partial Q_j} \mathrm{Re} \left( \sum_\alpha \frac{\partial \ell_{\alpha,t}}{\partial P_i} \frac{\partial \ell_{\alpha,t}^*}{\partial P_j} \right) + \frac{\partial}{\partial P_j} \mathrm{Re} \left( -\sum_\alpha \frac{\partial \ell_{\alpha,t}}{\partial P_i} \frac{\partial \ell_{\alpha,t}^*}{\partial Q_j} \right) \right\} W_{t}^{(\rho)} \right] \\
        & \qquad + \sum_{i,j} \frac{\partial}{\partial P_i} \left[ \left\{ \frac{\partial}{\partial Q_j} \mathrm{Re} \left( -\sum_\alpha \frac{\partial \ell_{\alpha,t}^*}{\partial P_j} \frac{\partial \ell_{\alpha,t}}{\partial Q_i} \right) + \frac{\partial}{\partial P_j} \mathrm{Re} \left( \sum_\alpha \frac{\partial \ell_{\alpha,t}}{\partial Q_i} \frac{\partial \ell_{\alpha,t}^*}{\partial Q_j} \right) \right\} W_{t}^{(\rho)} \right] \ .
    \end{aligned} \label{temp4}
\end{align}
Here, by noting that
\begin{align}
     \frac{\partial}{\partial Q_j} \mathrm{Re} \left( \sum_\alpha \frac{\partial \ell_{\alpha,t}}{\partial P_i} \frac{\partial \ell_{\alpha,t}^*}{\partial P_j} \right) + \frac{\partial}{\partial P_j} \mathrm{Re} \left( -\sum_\alpha \frac{\partial \ell_{\alpha,t}}{\partial P_i} \frac{\partial \ell_{\alpha,t}^*}{\partial Q_j} \right) = \sum_\alpha \Re\left\{\pdv{\ell_{\alpha,t}}{P_i}, \ell_{\alpha,t}^*\right\}_p
\end{align}
and
\begin{align}
    \frac{\partial}{\partial Q_j} \mathrm{Re} \left( -\sum_\alpha \frac{\partial \ell_{\alpha,t}^*}{\partial P_j} \frac{\partial \ell_{\alpha,t}}{\partial Q_i} \right) + \frac{\partial}{\partial P_j} \mathrm{Re} \left( \sum_\alpha \frac{\partial \ell_{\alpha,t}}{\partial Q_i} \frac{\partial \ell_{\alpha,t}^*}{\partial Q_j} \right) = \sum_\alpha \Re \left\{\ell_{\alpha,t}^*, \pdv{\ell_{\alpha,t}}{Q_i}\right\}_p \ ,
\end{align}
Eq. \eqref{temp4} is simplified as
\begin{align}
    &\sum_{\mu,\nu}\pdv{}{x^{\mu}}\left(\pdv{D_{R}^{\mu\nu}}{x^{\nu}}W_{t}^{(\rho)}\right) \nonumber \\
    & = \sum_i\left[\pdv{}{Q_i}\left(\sum_\alpha \Re\left\{\pdv{\ell_{\alpha,t}}{P_i}, \ell_{\alpha,t}^*\right\}_p W_{t}^{(\rho)}\right) + \pdv{}{P_i}\left(\sum_\alpha \Re \left\{\ell_{\alpha,t}^*, \pdv{\ell_{\alpha,t}}{Q_i}\right\}_p W_{t}^{(\rho)}\right)\right] \ .
\end{align}
Therefore, Eq. \eqref{temp3} is rewritten as
\begin{align}
    &\sum_{\mu,\nu}D_{R}^{\mu\nu}\pdv{W_{t}^{(\rho)}}{x^{\mu}}{x^{\nu}}\nonumber\\
    &\begin{aligned}
        &= \sum_{\mu,\nu} \frac{\partial^2}{\partial x^{\mu} \partial x^{\nu}} \left(D_{R}^{\mu\nu} W_{t}^{(\rho)}\right) \\
        &\qquad-2\sum_i\left[\pdv{}{Q_i}\left(\sum_\alpha \Re\left\{\pdv{\ell_{\alpha,t}}{P_i}, \ell_{\alpha,t}^*\right\}_p W_{t}^{(\rho)}\right) + \pdv{}{P_i}\left(\sum_\alpha \Re \left\{\ell_{\alpha,t}^*, \pdv{\ell_{\alpha,t}}{Q_i}\right\}_p W_{t}^{(\rho)}\right)\right]\\
        &\qquad + \sum_{i,j} \sum_\alpha \Biggl[ 2 \frac{\partial^2 \ell_{\alpha,t}}{\partial Q_j \partial P_i} \frac{\partial^2 \ell_{\alpha,t}^*}{\partial Q_i \partial P_j} - \left( \frac{\partial^2 \ell_{\alpha,t}}{\partial P_i \partial P_j} \frac{\partial^2 \ell_{\alpha,t}^*}{\partial Q_i \partial Q_j} + \frac{\partial^2 \ell_{\alpha,t}^*}{\partial P_i \partial P_j} \frac{\partial^2 \ell_{\alpha,t}}{\partial Q_i \partial Q_j} \right) \Biggr] \ .
    \end{aligned}
\end{align}
By combining the above results, we rewrite the terms of the first order in $\hbar$ in
Eq. \eqref{eq:temp70} as
\begin{align}
    &\begin{aligned}
    &=  \frac{\hbar}{2} \left[ \sum_i \frac{\partial}{\partial Q_i} \left(\sum_\alpha \mathrm{Re}\left\{ \ell_{\alpha,t}, \frac{\partial \ell_{\alpha,t}^*}{\partial P_i} \right\}_p W_{t}^{(\rho)} \right) + \sum_i \frac{\partial }{\partial P_i}\left(\sum_\alpha \mathrm{Re} \left\{ \frac{\partial \ell_{\alpha,t}^*}{\partial Q_i}, \ell_{\alpha,t} \right\}_p W_{t}^{(\rho)} \right)  \right]\\
    &\qquad + \frac{\hbar}{2} \sum_{\mu,\nu} \frac{\partial^2}{\partial x^{\mu} \partial x^{\nu}}\left(D_{R}^{\mu\nu}(X,t) W_{t}^{(\rho)}\right)
    \end{aligned}\label{eq:temp90}
\end{align}
Therefore, Eq. \eqref{eq:temp70} reduces to
\begin{align}
    &\left[\sum_\alpha\left(\hat{L}_{\alpha,t}\oprho_{t}\hat{L}_{\alpha,t}^{\dagger}-\frac{1}{2}\left[\hat{L}_{\alpha,t}^{\dagger}\hat{L}_{\alpha,t},\oprho_{t}\right]_{+}\right)\right]_{W}\nonumber\\
    &\begin{aligned}
    &\qquad= \sum_i\left[ \frac{\partial}{\partial Q_i} \left\{ \sum_\alpha \left(- \mathrm{Im} \left( \ell_{\alpha,t} \frac{\partial \ell_{\alpha,t}^*}{\partial P_i} \right) + \frac{\hbar}{2}\mathrm{Re}\left\{ \ell_{\alpha,t}, \frac{\partial \ell_{\alpha,t}^*}{\partial P_i} \right\}_p  \right) W_{t}^{(\rho)} \right\} \right. \\
    &\qquad\hspace{3em}\left. + \frac{\partial }{\partial P_i}\left\{ \sum_\alpha \left( - \mathrm{Im} \left( \frac{\partial \ell_{\alpha,t}}{\partial Q_i} \ell_{\alpha,t}^* \right) + \frac{\hbar}{2}\mathrm{Re} \left\{ \frac{\partial \ell_{\alpha,t}^*}{\partial Q_i}, \ell_{\alpha,t} \right\}_p \right) W_{t}^{(\rho)}\right\}  \right]\\
    &\qquad\quad + \frac{\hbar}{2} \sum_{\mu,\nu} \frac{\partial^2}{\partial x^{\mu} \partial x^{\nu}}\left(D_{R}^{\mu \nu}(X,t) W_{t}^{(\rho)}\right) + \order{\hbar^2} \ .
    \end{aligned} \label{temp5}
\end{align}
Finally, we calculate the first term in Eq. \eqref{eq:lindblad} as
\begin{align}
    \left[-\frac{i}{\hbar}\left[\hat{H}_{t},\oprho_{t}\right]_{-}\right]_{W} &= \left\{H_{t},W_{t}^{(\rho)}\right\}_{p} + \order{\hbar^3} \nonumber\\
    &=\sum_{i}\left[\pdv{H_t}{Q_i}\pdv{W_{t}^{(\rho)}}{P_i}-\pdv{H_t}{P_i}\pdv{W_{t}^{(\rho)}}{Q_i}\right] + \order{\hbar^3} \nonumber\\
    &=\sum_i\left[\pdv{}{P_i}\qty(\pdv{H_t}{Q_i}W_{t}^{(\rho)})-\pdv{}{Q_i}\qty(\pdv{H_t}{P_i}W_{t}^{(\rho)})\right] + \order{\hbar^3} \ . \label{eq:first term of Lindblad eq}
\end{align}
Thus, combining Eqs. \eqref{temp5} and \eqref{eq:first term of Lindblad eq} and setting
$G^{\mu\nu} = \hbar D_R^{\mu\nu}$ yields Eq. \eqref{eq:Wigner tran of Lindblad eq} in section 3.

\subsection{$\hbar$ expansion of reversed Lindblad equation}
\label{appendix:hbar expansion of reversal Lindblad eq}
In this section, we perform the Wigner transformation of Eq. \eqref{eq:reversal Lindblad eq} and expand the Wigner transform
up to $\mathcal{O}(\hbar)$.
We denote the Wigner transforms of the reversed Hamiltonian
$\hat{\tilde{H}}_t$
\eqref{eq:reversal hamiltonian},
the reversed Lindblad operators $\hat{\tilde{L}}_{\alpha,t}$ \eqref{eq:reversal Lindblad operator}
and the operator $\hat{G}$ \eqref{G} by $\tilde{H}_t$, $\tilde{\ell}_{\alpha,t}/\sqrt{\hbar}$ and $G_W$, respectively.

The reversed Hamiltonian $\hat{\tilde{H}}_t$ can be rewritten as
\begin{align}
    \hat{\tilde{H}}_t = - \hat{H}_t + \frac{1}{2}\left[ \left[ \hat{H}_t, \hat{G} \right]_{-},\hat{G}^{-1} \right]_{-} - \frac{i\hbar}{2}\left[\dot{\hat{G}},\hat{G}^{-1}\right]_{-} + \frac{i}{4\hbar} \sum_\alpha \left[\left[\hat{L}_{\alpha,t}^{\dagger}\hat{L}_{\alpha,t},\hat{G}\right]_{-},\hat{G}^{-1}\right]_{+} \ .
    \label{temp6}
\end{align}
Here, the Wigner transforms of the second and third terms on the right-hand side of Eq. \eqref{temp6} read
\begin{align}
    \left[\frac{1}{2}\left[ \left[ \hat{H}_t, \hat{G} \right]_{-},\hat{G}^{-1} \right]_{-}\right]_{W} &= \frac{1}{2}\left(i\hbar\qty{\left[H_t, G_W\right]_{-}, G_W^{-1}}_p + \order{\hbar^3}\right) \nonumber\\
    &=\frac{i\hbar}{2}\qty{i\hbar\qty{H_t,G_W}_p, G_W^{-1}}_p + \order{\hbar^3} \nonumber\\
    &=\order{\hbar^2} \ ,
\end{align}
and
\begin{align}
    \left[ \frac{i\hbar}{2}\left[\hat{\dot{G}}, \hat{G}^{-1}\right]_{-} \right]_{W} = \frac{i\hbar}{2}\left(i\hbar\qty{\dot{G}_W,G_W^{-1}}_p + \order{\hbar^3}\right) = \order{\hbar^2} \ .
\end{align}
Thus, the Wigner transform of the reversed Hamiltonian is obtained as
\begin{align}
    \tilde{H}_t = \left[\hat{\tilde{H}}_t\right]_W = -H_t + \frac{i}{4} \sum_\alpha \left[\left[\ell_{\alpha,t}^* \star  \ell_{\alpha,t},G_W\right]_{-},G_W^{-1}\right]_{+} + \order{\hbar^2} \ .  \label{temp7}
\end{align}
The second term reduces to
\begin{align}
    &\frac{i}{4} \sum_\alpha \left[\left[\ell_{\alpha,t}^* \star  \ell_{\alpha,t},G_W\right]_{-},G_W^{-1}\right]_{+}
    = \frac{i}{2} \sum_\alpha \left[i\hbar\qty{\abs{\ell_{\alpha,t}}^2,G_W}_p G_W^{-1} \right] + \order{\hbar^2} \ .
\end{align}
To the zeroth-order in $\hbar$, we can regard as $G_{W} = (W_{t}^{(\gamma)})^{1/2}$ and $ G_{W}^{-1} = (W_{t}^{(\gamma)})^{-1/2}$ so that
\begin{align}
    \pdv{G_{W}}{x^{\mu}}G_W^{-1} = \pdv{(W_{t}^{(\gamma)})^{1/2}}{x^{\mu}} \left(W_{t}^{(\gamma)}\right)^{-1/2}  = \frac{1}{2}\qty(\frac{1}{W_{t}^{(\gamma)}}\pdv{W_{t}^{(\gamma)}}{x^{\mu}}) \ . \label{eq:temp 100}
\end{align}
Thus, Eq. \eqref{temp7} leads to
\begin{align}
    &\frac{i}{4} \sum_\alpha \left[\left[\ell_{\alpha,t}^* \star  \ell_{\alpha,t},G_W\right]_{-},G_W^{-1}\right]_{+} \nonumber \\
    & = \frac{i}{2} \sum_\alpha \left[i\hbar\qty{\abs{\ell_{\alpha,t}}^2,G_W}_p G_W^{-1} \right] + \order{\hbar^2} \nonumber\\
    & = -\frac{\hbar}{2} \sum_\alpha \sum_i \left[\qty{\pdv{\abs{\ell_{\alpha,t}}^2}{Q_i}\pdv{G_W}{P_i} - \pdv{\abs{\ell_{\alpha,t}}^2}{P_i}\pdv{G_W}{Q_i}} G_W^{-1} \right] + \order{\hbar^2} \nonumber\\
    & = -\frac{\hbar}{4} \sum_\alpha \sum_i \left[\pdv{\abs{\ell_{\alpha,t}}^2}{Q_i}\frac{1}{W_{t}^{(\gamma)}}\pdv{W_{t}^{(\gamma)}}{P^{i}} - \pdv{\abs{\ell_{\alpha,t}}^2}{P_i}\frac{1}{W_{t}^{(\gamma)}}\pdv{W_{t}^{(\gamma)}}{Q^{i}} \right] + \order{\hbar^2} \nonumber\\
    &=-\frac{\hbar}{4} \sum_\alpha \sum_i \left[\Re{\ell_{\alpha,t}^*\pdv{\ell_{\alpha,t}}{Q_{i}}}V_{P_i} - \Re{\ell_{\alpha,t}^*\pdv{\ell_{\alpha,t}}{P_{i}}}V_{Q_i} \right] + \order{\hbar^2} \ .
\end{align}
In the last line, we have defined
\begin{align}
    V_{Q_i} \coloneqq \frac{1}{W_{t}^{(\gamma)}}\pdv{W_{t}^{(\gamma)}}{Q_{i}},\quad V_{P_i} \coloneqq \frac{1}{W_{t}^{(\gamma)}}\pdv{W_{t}^{(\gamma)}}{P_{i}} \ .
\end{align}
From the above calculation, the Wigner transform of the reversed Hamiltonian is obtained as
\begin{align}
    \tilde{H}_t = -H_t -\frac{\hbar}{4} \sum_\alpha \sum_i \left[\Re{\ell_{\alpha,t}^*\pdv{\ell_{\alpha,t}}{Q_{i}}}V_{P_i} - \Re{\ell_{\alpha,t}^*\pdv{\ell_{\alpha,t}}{P_{i}}}V_{Q_i} \right] + \order{\hbar^2} \ .
\end{align}
Hence, to the leading order in $\hbar$, the reversed Hamiltonian acquires an additional drift proportional
to the gradient of the reference Wigner function.

Next, we consider the Wigner transform of the reversed Lindblad operator \eqref{eq:reversal Lindblad operator}.
By using an equality
\begin{align}
\hat{G}\hat{L}_{\alpha,t}^{\dagger}\hat{G}^{-1} = \hat{L}_{\alpha,t}^{\dagger} - \hat{G}\left[\hat{G}^{-1},\hat{L}_{\alpha,t}^{\dagger}\right]_{-} \ ,
\end{align}
we obtain
\begin{align}
    \tilde{\ell}_{\alpha,t}
    & = \ell_{\alpha,t}^* - G_{W} \star \left[G_{W}^{-1}, \ell_{\alpha,t}^*\right]_{-}  \nonumber\\
    & = \ell_{\alpha,t}^* - G_{W} \star \qty(i\hbar\qty{G_{W}^{-1},\ell_{\alpha,t}^*}_p + \order{\hbar^3}) \nonumber\\
    & = \ell_{\alpha,t}^* - i\hbar G_{W} \qty{G_{W}^{-1},\ell_{\alpha,t}^*}_p + \order{\hbar^2} \nonumber\\
    & = \ell_{\alpha,t}^* + \frac{i\hbar}{2} \sum_{i} \qty(\pdv{\ell_{\alpha,t}^*}{P_{i}} V_{Q_{i}} - \pdv{\ell_{\alpha,t}^*}{Q_{i}} V_{P_{i}}) + \order{\hbar^2} \ .\label{eq:temp 91}
\end{align}
Here, we have used Eq. \eqref{eq:temp 100}. In a similar way, from
\begin{align}
\sqrt{\hbar}\hat{G}^{-1}\hat{L}_{\alpha,t}\hat{G} = \hat{\ell}_{\alpha,t} - \left[\hat{\ell}_{\alpha,t}, \hat{G}^{-1}\right]_{-}\hat{G} \ ,
\end{align}
we obtain
\begin{align}
    \tilde{\ell}_{\alpha,t}^* = \ell_{\alpha,t} - \frac{i\hbar}{2}\sum_{i}\qty(\pdv{\ell_{\alpha,t}}{P_{i}} V_{Q_{i}} - \pdv{\ell_{\alpha,t}}{Q_{i}} V_{P_{i}}) + \order{\hbar^2} \ .\label{eq:temp 92}
\end{align}

Finally, we substitute the above results into Eq. \eqref{Wigner tran of inverse Lindblad eq}.
First, we consider the diffusion term. Since
\begin{align}
    \tilde{G}^{\mu\nu} = \left(G^{\mu\nu}\right)^{*} + \order{\hbar^2} = G^{\nu\mu} + \order{\hbar^2} \ ,
\end{align}
the diffusion term reduces to
\begin{align}
    \frac{1}{2}\sum_{\mu,\nu} \frac{\partial^2}{\partial x^{\mu} \partial x^{\nu}} \left( \tilde{G}^{\mu\nu}(\bm{x},t) W_{t}^{(\rho)} \right) = \frac{1}{2}\sum_{\mu,\nu} \frac{\partial^2}{\partial x^{\mu} \partial x^{\nu}} \left( G^{\mu\nu}(\bm{x},t) W_{t}^{(\rho)} \right) + \order{\hbar^2} \ .
\end{align}
Next, we consider the drift term. Some calculations yield
\begin{align}
    &\begin{aligned}
        \frac{\partial \tilde{H}_t}{\partial P_i} = - \frac{\partial H_t}{\partial P_i} - \frac{\hbar}{2} \sum_\alpha \sum_j \mathrm{Re} \Biggl[ &\frac{\partial \ell_{\alpha,t}^*}{\partial P_i} \frac{\partial \ell_{\alpha,t}}{\partial Q_j} V_{P_j} + \ell_{\alpha,t}^* \frac{\partial^2 \ell_{\alpha,t}}{\partial P_i \partial Q_j} V_{P_j} + \ell_{\alpha,t}^* \frac{\partial \ell_{\alpha,t}}{\partial Q_j} \frac{\partial V_{P_j}}{\partial P_i}  \nonumber\\
        & - \frac{\partial \ell_{\alpha,t}^*}{\partial P_i} \frac{\partial \ell_{\alpha,t}}{\partial P_j} V_{Q_{j}} - \ell_{\alpha,t}^* \frac{\partial^2 \ell_{\alpha,t}}{\partial P_i \partial P_j} V_{Q_{j}} - \ell_{\alpha,t}^* \frac{\partial \ell_{\alpha,t}}{\partial P_j} \frac{\partial V_{Q_{j}}}{\partial P_i} \Biggr]
    \end{aligned}
    \\
    &\begin{aligned}
        = &- \frac{\partial H_t}{\partial P_i} + \frac{\hbar}{2} \sum_\alpha\sum_j \mathrm{Re} \Biggl[ \frac{1}{W_t^{(\gamma)}} \left\{  \frac{\partial}{\partial P_j} \left( - \frac{\partial \ell_{\alpha,t}^*}{\partial P_i} \frac{\partial \ell_{\alpha,t}}{\partial Q_j} W_t^{(\gamma)} \right) + \frac{\partial}{\partial Q_j} \left( \frac{\partial \ell_{\alpha,t}^*}{\partial P_i} \frac{\partial \ell_{\alpha,t}}{\partial P_j} W_t^{(\gamma)} \right) \right\}\Biggr] \\
        &- \frac{\hbar}{2} \sum_\alpha \sum_j \mathrm{Re} \Biggl[\ell_{\alpha,t}^* \frac{\partial^2 \ell_{\alpha,t}}{\partial P_i \partial Q_j} V_{P_j} + \ell_{\alpha,t}^* \frac{\partial \ell_{\alpha,t}}{\partial Q_j} \frac{\partial V_{P_j}}{\partial P_i} - \ell_{\alpha,t}^* \frac{\partial^2 \ell_{\alpha,t}}{\partial P_i \partial P_j} V_{Q_j} - \ell_{\alpha,t}^* \frac{\partial \ell_{\alpha,t}}{\partial P_j} \frac{\partial V_{Q_j}}{\partial P_i} \Biggr]\\
        & -\frac{\hbar}{2}\sum_{\alpha}\Re\qty{\pdv{\ell_{\alpha,t}}{P_i},\ell_{\alpha,t}^*}_p  \ ,
    \end{aligned}
\end{align}
\begin{align}
    &\begin{aligned}
        \Im\qty(\tilde{\ell}_{\alpha,t}\pdv{\tilde{\ell}_{\alpha,t}^*}{P_{i}}) &= \Im\qty(\ell_{\alpha,t}^*\pdv{\ell_{\alpha,t}}{P_{i}}) + \frac{\hbar}{2}\sum_j\Re{\pdv{\ell_{\alpha,t}}{P_{i}}\qty(\pdv{\ell_{\alpha,t}^*}{P_j}V_{Q_{j}}-\pdv{\ell_{\alpha,t}^*}{Q_{j}}V_{P_j})}\\
        &\qquad-\frac{\hbar}{2}\sum_j\Re{\ell_{\alpha,t}^*\pdv{}{P_{i}}\qty(\pdv{\ell_{\alpha,t}}{P_j}V_{Q_{j}}-\pdv{\ell_{\alpha,t}}{Q_{j}}V_{P_j})} + \order{\hbar^2}
    \end{aligned} \nonumber\\
    &\begin{aligned}
        &= \mathrm{Im} \left( \ell_{\alpha,t}^* \frac{\partial \ell_{\alpha,t}}{\partial P_i} \right) + \frac{\hbar}{2} \mathrm{Re} \Biggl[ \frac{1}{W_t^{(\gamma)}} \left\{ \frac{\partial}{\partial Q_j} \left( \frac{\partial \ell_{\alpha,t}}{\partial P_i} \frac{\partial \ell_{\alpha,t}^*}{\partial P_j} W_t^{(\gamma)} \right) + \frac{\partial}{\partial P_j} \left( -\frac{\partial \ell_{\alpha,t}}{\partial P_i} \frac{\partial \ell_{\alpha,t}^*}{\partial Q_j} W_t^{(\gamma)} \right) \right\}\Biggr] \\
        &+ \frac{\hbar}{2} \sum_\alpha \sum_j \mathrm{Re} \Biggl[\ell_{\alpha,t}^* \frac{\partial^2 \ell_{\alpha,t}}{\partial P_i \partial Q_j} V_{P_j} + \ell_{\alpha,t}^* \frac{\partial \ell_{\alpha,t}}{\partial Q_j} \frac{\partial V_{P_j}}{\partial P_i} - \ell_{\alpha,t}^* \frac{\partial^2 \ell_{\alpha,t}}{\partial P_i \partial P_j} V_{Q_j} - \ell_{\alpha,t}^* \frac{\partial \ell_{\alpha,t}}{\partial P_j} \frac{\partial V_{Q_j}}{\partial P_i} \Biggr]\\
        & -\frac{\hbar}{2}\sum_{\alpha}\Re\qty{\pdv{\ell_{\alpha,t}}{P_i},\ell_{\alpha,t}^*}_p \ ,
        \end{aligned}
\end{align}
and
\begin{align}
    \frac{\hbar}{2}\Re\qty{\tilde{\ell}_{\alpha,t},\pdv{\tilde{\ell}_{\alpha,t}^{*}}{P_{i}}}_{p} = \frac{\hbar}{2}\Re\qty{\ell^*_{\alpha,t}, \pdv{\ell_{\alpha,t}}{P_{i}}}_p + \order{\hbar^2} \ .
\end{align}
By substituting the above results into Eq. \eqref{drift 1},
we obtain for $\mu=i=1,\ldots,N$
\begin{align}
    &\begin{aligned}
    \tilde{f}^{\mu}(\bm{x},t) =& -\pdv{H_t}{P_i} - \sum_\alpha \Im\qty(\ell_{\alpha,t}\pdv{\ell_{\alpha,t}^*}{P_i}) + \sum_\alpha \frac{\hbar}{2}\Re\qty{\ell_{\alpha,t},\pdv{\ell_{\alpha,t}^*}{P_i}}  \\
    &+ \hbar\sum_\alpha\sum_j\mathrm{Re} \Biggl[ \frac{1}{W_t^{(\gamma)}} \left\{ \frac{\partial}{\partial Q_j} \left( \frac{\partial \ell_{\alpha,t}}{\partial P_i} \frac{\partial \ell_{\alpha,t}^*}{\partial P_j} W_t^{(\gamma)} \right) + \frac{\partial}{\partial P_j} \left( -\frac{\partial \ell_{\alpha,t}}{\partial P_i} \frac{\partial \ell_{\alpha,t}^*}{\partial Q_j} W_t^{(\gamma)} \right)\right\}\Biggr] \nonumber\\
    &+\order{\hbar^2}
    \end{aligned}\\
    &\begin{aligned}
        \hphantom{\tilde{f}^{i}(\bm{x},t)} = - f^{\mu}(\bm{x},t) + \frac{1}{W_{t}^{(\gamma)}}\sum_{\nu}\pdv{}{x^{\nu}}\qty(G^{\mu\nu}W_{t}^{(\gamma)}) + \order{\hbar^2} \ .
    \end{aligned}
\end{align}
In a similar way,
we obtain for $\mu=i+N=N+1,\cdots, 2N$
\begin{align}
    &\begin{aligned}
    \tilde{f}^{\mu}(\bm{x},t) =& \pdv{H_t}{Q_{i}} - \sum_{\alpha} \Im\qty(\pdv{\ell_{\alpha,t}}{Q_{i}}) + \sum_\alpha\frac{\hbar}{2}\Re\qty{\pdv{\ell_{\alpha,t}}{Q_{i}},\ell_{\alpha,t}^*}_p\\
    & + \hbar \sum_\alpha \sum_j \Re\qty[\frac{1}{W_{t}^{(\gamma)}}\qty{\pdv{}{Q_j}\qty(-\pdv{\ell_{\alpha,t}}{P_j}\pdv{\ell_{\alpha,t}^*}{Q_i}W_{t}^{(\gamma)}) + \pdv{}{P_j}\qty(\pdv{\ell_{\alpha,t}}{Q_i}\pdv{\ell_{\alpha,t}^*}{Q_j}W_{t}^{(\gamma)})}]\\
    & + \order{\hbar^2}
    \end{aligned} \nonumber\\
    &\begin{aligned}
        \hphantom{\tilde{f}^{\mu}(\bm{x},t)} = -f^{\mu}(\bm{x},t) + \frac{1}{W_{t}^{(\gamma)}}\sum_{\nu}\pdv{}{x^{\nu}}\qty(G^{\mu\nu}W_{t}^{(\gamma)}) + \order{\hbar^2} \ .
    \end{aligned}
\end{align}
Thus, we obtain Eq. \eqref{eq:1st order of Wigner tran of inverse Lindblad eq} in section 3.

\section{Example of WKB solution}
In this appendix, we construct a concrete WKB solution for the following model:
\begin{gather}
        H(x)=\frac{P^2}{2m} + \frac{1}{2}kQ^2 \ ,\\
        \ell_{\alpha}(x) = \sigma_\alpha P + \theta_\alpha Q \ ,
\end{gather}
where $\sigma_\alpha$ and $\theta_\alpha$ are complex numbers.
We consider the forward process in appendix B.1 and the reverse process in appendix B.2.

\subsection{Forward process}
In this model, $K_\mu$ are given by
\begin{align}
    K_p &= \Im\qty(\sum_\alpha\qty(\sigma_\alpha P + \theta_\alpha Q)\sigma_\alpha^*) = \Im\qty(\sum_\alpha\theta_\alpha\sigma_\alpha^*)Q \ ,\\
    K_q &= \Im\qty(\sum_\alpha\qty(\sigma_\alpha P + \theta_\alpha Q)\theta_\alpha^*) = -\Im\qty(\sum_\alpha\theta_\alpha\sigma_\alpha^*)P \ .
\end{align}
On the other hand, since $J_\mu$ contains second derivatives of $\ell_\alpha$, in this model,
we have
\begin{align}
    J_\mu &= 0 \ .
\end{align}
Furthermore, by noting that the derivatives of the Hamiltonian are
\begin{align}
     \pdv{Q} H = kQ \ , \quad \pdv{P} H = \frac{P}{m} \ ,
\end{align}
we obtain
\begin{align}
    A = \frac{P}{m} + fQ \  \;\;\;
    B = fP-kQ \ , \;\;\;
    \Gamma = 2f \ ,
\end{align}
where we have defined $f=\Im\qty(\sum_\alpha\theta_\alpha\sigma_\alpha^*)$. Hence, the equations for the characteristic curves are
\begin{align}
    \dot{Q} &= \frac{P}{m} + f Q  \ , \\
    \dot{P} &= f P - k Q \ .
\end{align}
These coupled equations can be solved to give the characteristic curve:
\begin{align}
    Q(t) &= e^{ft}\qty(\cos\omega t\,Q_0-\frac{1}{m\omega}\sin\omega t\,P_0  ) \ ,
    \label{Q}\\
    P(t) &= e^{ft}\qty( m\omega\sin \omega t\, Q_0+\cos\omega t\, P_0 ) \ ,
    \label{P}
\end{align}
where $Q_0=Q(0),\,P_0=P(0)$ and $\omega = \sqrt{k/m}$.
Hence,
we can construct the map $\Phi_t$ and its
inverse $\Phi^{-1}_t$ from Eqs. \eqref{Q} and \eqref{P}
and find that
the zeroth-order solution is
\begin{align}
    W_{t,0}^{(\rho)} = w_0(x,t)e^{-2ft}\label{eq:harmonic_oscillator_0th} \ ,
\end{align}
where
\begin{align}
    w_0(x,t) = W_{0,0}^{(\rho)}(e^{-ft}(\cos\,\omega t\,Q+\sin\,\omega t/m\omega\,P ),e^{-ft}(- m\omega \sin\,\omega t\,Q+\cos\,\omega t\,P )) \ . \label{eq:harmonic_oscillator_0th_initial}
\end{align}

The first-order solution is obtained as follows. By using
Eqs. \eqref{eq:harmonic_oscillator_0th} and \eqref{eq:harmonic_oscillator_0th_initial}, we
calculate $\Delta_0$. Since $\ell_\alpha$ is linear in $x$, $D^{\mu\nu}$ is
independent of $x$. Then, we obtain
\begin{align}
    \Delta_0(x,t) = \frac{1}{2}D^{\mu\nu}\ppdiff{}{x^\mu}{x^\nu} W_{t,0}^{(\rho)} \ .
\end{align}
Here $D^{\mu\nu}$ are given by
\begin{align}
    D^{QQ}&=\sum_\alpha\abs{\sigma_\alpha}^2  \ ,\\
    D^{QP}&=D^{PQ}= -\sum_\alpha\Re\qty(\theta_\alpha\sigma_\alpha^*) \ , \\
    D^{PP}&=\sum_\alpha\abs{\theta_\alpha}^2 \ .
\end{align}
As for $\ppdiff{}{x^\mu}{x^\nu} W_{t,0}^{(\rho)}$, we have
\begin{align}
    \ppdiff{}{x^\mu}{x^\nu} W_{t,0}^{(\rho)} &= e^{-2ft}\ppdiff{}{x^\mu}{x^\nu} w_0(x,t) \nonumber \\
    &=e^{-2ft}\ppdiff{}{x^\mu}{x^\nu} W_{0,0}^{(\rho)}(e^{-ft}(\cos\,\omega t\,Q +\sin\,\omega t/m\omega\,P), e^{-ft}( - m\omega \sin\,\omega t\,Q + \cos\,\omega t\,P))
\end{align}
so that
\begin{align}
    &\pdv[2]{Q} W_{t,0}^{(\rho)} \nonumber\\
    &=e^{-2ft}\left.\qty(\qty(\pdv{Q_0}{Q})^2\pdv[2]{Q_0}W_{0,0}^{(\rho)}(x_0) + 2\pdv{Q_0}{Q}\pdv{P_0}{Q}\ppdiff{}{Q_0}{P_0}W_{0,0}^{(\rho)}(x_0) + \qty(\pdv{P_0}{Q})^2\pdv[2]{P_0}W_{0,0}^{(\rho)}(x_0))\right|_{x_0=\Phi_t^{-1}(x)} \nonumber \\
    &=\left.e^{-4ft}\qty(\cos^2\omega t\pdv[2]{Q_0}W_{0,0}^{(\rho)}(x_0) - m\omega\sin2\omega t\ppdiff{}{Q_0}{P_0}W_{0,0}^{(\rho)}(x_0)+m^2\omega^2\sin^2\omega t\, \pdv[2]{P_0}W_{0,0}^{(\rho)}(x_0))\right|_{x_0=\Phi_t^{-1}(x)} \ , \\
     &\pdv[2]{P} W_{t,0}^{(\rho)} \nonumber\\
    &=\left.e^{-4ft}\qty(\frac{1}{m^2\omega^2}\cos^2\omega t\pdv[2]{Q_0}W_{0,0}^{(\rho)}(x_0)+\frac{1}{m\omega}\sin2\omega t\ppdiff{}{Q_0}{P_0}W_{0,0}^{(\rho)}(x_0)+\cos^2\omega t\, \pdv[2]{P_0}W_{0,0}^{(\rho)}(x_0) )\right|_{x_0=\Phi_t^{-1}(x)} \ ,\\
     &\ppdiff{}{Q}{P} W_{t,0}^{(\rho)} \nonumber\\
    &=\left.e^{-4ft}\qty(\frac{1}{2m\omega}\sin2\omega t\pdv[2]{Q_0}W_{0,0}^{(\rho)}(x_0) +\cos2\omega t\ppdiff{}{Q_0}{P_0}W_{0,0}^{(\rho)}(x_0)-\frac{m\omega}{2}\sin2\omega t\, \pdv[2]{P_0}W_{0,0}^{(\rho)}(x_0))\right|_{x_0=\Phi_t^{-1}(x)} \ .
\end{align}
Once $W_{0,0}^{(\rho)}(x)$ is given, $\Delta_0$ can be determined completely by substituting it into these formulae. Substituting that result into Eq. \eqref{eq:solution_for_W1} and performing the time integral yields the solution.

\subsection{Reverse process}
In this model, $\bar{A}$, $\bar{B}$ and $\bar{\Gamma}$ are given by
\begin{align}
    \bar{A} &= \frac{1}{m}\bar{P} - f\bar{Q} \ ,\\
    \bar{B} &= f\bar{P} + k\bar{Q} \ ,\\
    \bar{\Gamma} &= -2f \ .
\end{align}
Therefore, the reverse characteristic equations for this model are
\begin{align}
    \whitedot{\bar{Q}}(\bar{t}) &= \frac{1}{m}\bar{P}(\bar{t}) - f \bar{Q}(\bar{t}) \ , \\
    \whitedot{\bar{P}}(\bar{t}) &= -f\bar{P}(\bar{t}) - k \bar{Q}(\bar{t}) \ .
\end{align}
We can solve these coupled equations as
\begin{align}
    \bar{Q}(\bar{t}) &= e^{-f\bar{t}}\qty(-\frac{1}{m\omega}\sin{\omega\bar{t}}\,\bar{P}_0 + \cos{\omega\bar{t}}\,\bar{Q}_0) \ , \label{eq:backward_characteristic_curve_q}  \\
    \bar{P}(\bar{t}) &= e^{-f\bar{t}}\qty(\cos{\omega\bar{t}}\,\bar{P}_0 + m\omega\sin{\omega\bar{t}}\,\bar{Q}_0) \ , \label{eq:backward_characteristic_curve_p}
\end{align}
where $\bar{Q}_0=\bar{Q}(\bar{t}=0),\,\bar{P}_0=\bar{P}(\bar{t}=0)$ and $\omega = \sqrt{k/m}$.
By noting that, for Eqs. \eqref{eq:backward_characteristic_curve_p} and \eqref{eq:backward_characteristic_curve_q},
\begin{align}
    \bar{P}(\bar{t}) = \left.-P(t)\right|_{t=T-\bar{t}},\quad \bar{Q}(\bar{t}) = \left.Q(t)\right|_{t=T-\bar{t}} \ ,
\end{align}
we have
\begin{align}
    \bar{Q}_0 &= \bar{Q}(\bar{t}=0) = Q(T) = Q_T   \ , \\
    \bar{P}_0 &= \bar{P}(\bar{t}=0) = -P(T) = -P_T \ .
\end{align}
Using these, we obtain
\begin{align}
    Q(t) &= e^{-f(T-t)}\qty(-\frac{1}{m\omega}\sin(\omega(T-t))\,P_T + \cos(\omega(T-t))\,Q_T)  \ , \label{eq:backward_characteristic_curve_q2} \\
    -P(t) &= e^{-f(T-t)}\qty(-\cos(\omega(T-t))\,P_T + m\omega\sin(\omega(T-t))\,Q_T) \ . \label{eq:backward_characteristic_curve_p2}
\end{align}
From the above, we can construct the map
$\Psi_{\bar{t}}$ and its inverse map and
find that the zeroth-order solution of the reverse process is
\begin{align}
    \bar{W}^{(\rho)}_0(\bar{Q},\bar{P},\bar{t}) = \bar{w}_0(\bar{Q},\bar{P},\bar{t})e^{-f\bar{t}} \ ,
\end{align}
where
\begin{align}
    \bar{w}_0(\bar{Q},\bar{P},\bar{t}) &= \bar{W}^{(\rho)}_{0,0}(\Psi_{\bar{t}}^{-1}(\bar{Q},\bar{P}))\nonumber \\
    &=\bar{W}^{(\rho)}_{0,0}(e^{f\bar{t}}( - \cos\omega\bar{t}\,\bar{Q}+\sin\omega\bar{t}/m\omega\,\bar{P}),e^{f\bar{t}}(-m\omega\sin\omega\bar{t}\,\bar{Q}+\cos\omega\bar{t}\,\bar{P})) \ .
\end{align}

\bibliographystyle{utphys}
\bibliography{reference_Wigner}

\begin{thebibliography}{10}
\bibitem{footnotegammat}
Here $\hat{\gamma}_t$ is a reference state following Lindblad equation
  \eqref{eq:lindblad}.
\bibitem{footnoteinverseWigner}
The inverse of the Wigner transformation is given by
  $\hat{O}(\hat{\bm{q}},\hat{\bm{p}}) =
  \int\frac{d^{N}Qd^{N}P}{(2\pi\hbar)^N}O(\bm{Q},\bm{P})\hat{\Delta}(\bm{Q},\bm{P})
  \;\;$with$ \;\;\hat{\Delta}(\bm{Q},\bm{P}) = \int\frac{d^{N}\xi
  d^{N}\eta}{(2\pi\hbar)^N}e^{\frac{i}{\hbar}(\bm{Q}-\hat{\bm{q}})\cdot\bm{\xi}+\frac{i}{\hbar}(\bm{P}-\hat{\bm{p}})\cdot\bm{\eta}}
  $, which defines the Weyl ordering.
\bibitem{footnoteLindbladeq}
If $G$ has a zero eigenvalue, the system behaves deterministically in that
  direction of its eigen-vector.
\bibitem{footnoteWKB}
It is possible that one must consider a problem of connection around a turning
  point as in the WKB analysis of Schr\"odinger equation. In that sense, our
  solutions are valid only for regions free from such a problem.
\end{thebibliography}

\end{document}